\documentclass[aps,prr,longbibliography,twocolumn,superscriptaddress,amsmath,amssymb,nofootinbib]{revtex4-2}
\usepackage{graphicx,enumitem}
\usepackage{standalone}
\usepackage{multirow,dcolumn}
\usepackage{txfonts}
\usepackage{hyperref}
\usepackage{soul,cancel}
\usepackage{epsf,amsmath,amssymb,verbatim,color,multirow,pifont,graphicx,cleveref,tabularx}
\usepackage[dvipsnames]{xcolor}
\usepackage{rotating}
\usepackage{amsmath,amsfonts,amssymb,bm,cancel}

\usepackage{tikz}

\begin{document}
\title{
Absorbing phase transition with a continuously varying exponent in a quantum contact process: a neural network approach}

\author{Minjae Jo}
\affiliation{CCSS, CTP and Department of Physics and Astronomy, 
Seoul National University, Seoul 08826, Korea}

\author{Jongshin Lee}
\affiliation{CCSS, CTP and Department of Physics and Astronomy, 
Seoul National University, Seoul 08826, Korea}

\author{K. Choi}
\affiliation{CCSS, CTP and Department of Physics and Astronomy, 
	Seoul National University, Seoul 08826, Korea}

\author{B. Kahng}
\email{bkahng@snu.ac.kr}
\affiliation{CCSS, CTP and Department of Physics and Astronomy, 
Seoul National University, Seoul 08826, Korea}


\begin{abstract}
Phase transitions in dissipative quantum systems are intriguing because they are induced by the interplay between coherent quantum and incoherent classical fluctuations. Here, we investigate the crossover from a quantum to a classical absorbing phase transition arising in the quantum contact process (QCP). The Lindblad equation contains two parameters, $\omega$ and $\kappa$, which adjust the contributions of the quantum and classical effects, respectively. We find that in one dimension, when the QCP starts from a homogeneous state with all active sites, there exists a critical line in the region $0 \le \kappa < \kappa_*$ along which the exponent $\alpha$ (which is associated with the density of active sites) decreases continuously from a quantum to the classical directed percolation (DP) value. This behavior suggests that the quantum coherent effect remains to some extent near $\kappa=0$. 
 However, when the QCP in one dimension starts from a heterogeneous state with all inactive sites except for one active site, all the critical exponents have the classical DP values for $\kappa \ge 0$. In two dimensions, anomalous crossover behavior does not occur, and classical DP behavior appears in the entire region of $\kappa \ge 0$ regardless of the initial configuration. Neural network machine learning is used to identify the critical line and determine the correlation length exponent. Numerical simulations using the quantum jump Monte Carlo technique and tensor network method are performed to determine all the other critical exponents of the QCP.
\end{abstract}

\maketitle


\section{Introduction}
Quantum critical phenomena in nonequilibrium systems have attracted considerable attention recently \cite{carusotto2013,noh2016,carmichael2015,muller2012,baumann2010,baumann2011,bloch2005,fink2017,fink2018,fitzpatrick2017,espigares2017,helmrich2020,lee2013,jin2016,boite2013,klinder2015,zou2014,nagy2015,houck2012} with the development of experimental techniques in cold atomic physics such as the use of trapped ions~\cite{muller2012} and lattices of ultracold ions~\cite{baumann2010,baumann2011,bloch2005}; driven circuit quantum electrodynamics systems~\cite{fink2017}; and semiconductor microcavities~\cite{fink2018}. The quantum criticality in the equilibrium state may be perturbed by the external environment, and thus the combined system is left in a nonequilibrium state. Here, we are interested in dissipative phase transitions arising from competition between the coherent Hamiltonian dynamics and incoherent dissipation processes~\cite{sieberer2013,sieberer2014,lang2016,diehl2010,torre2010,torre2012,tauber2014,mitra2006,torre2013,rota2017,rota2019,verstraelen2020,marino2016}.
For these systems, questions arise as to whether the competition between quantum coherent and classical incoherent fluctuations produces another type of universal behavior~\cite{marino2016,lang2016} and the conditions under which they exhibit classical critical behavior in terms of the loss rates to the environment~\cite{mitra2006,torre2010,torre2012}.

\begin{figure}[t!]
\includegraphics[width=0.85\columnwidth]{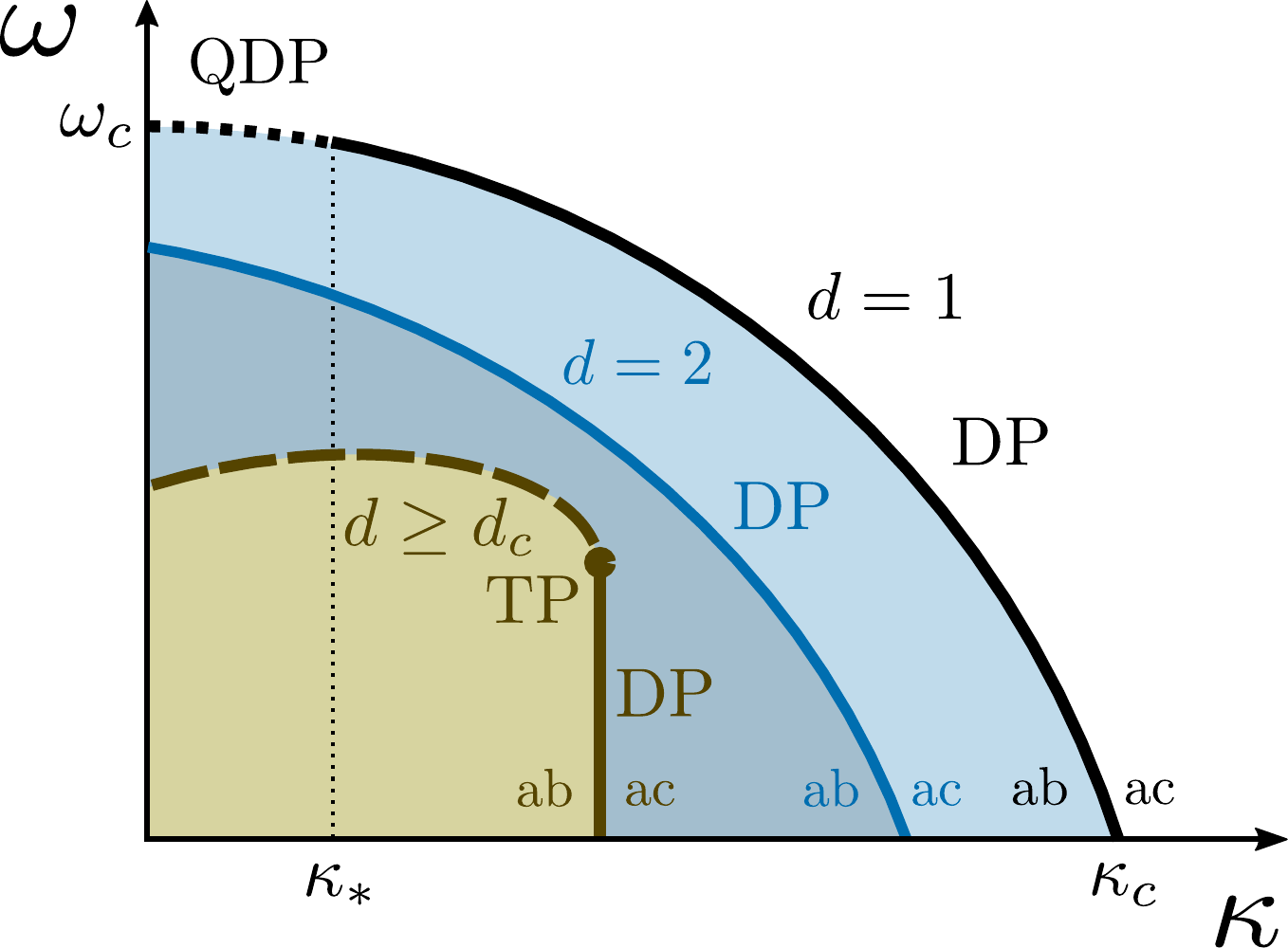}
\caption{Schematic phase diagram of the QCP in the parameter space ($\kappa$, $\omega$) in the mean-field limit (inside, $d\ge d_c$, where $d$ is the spatial dimension, and $d_c=3$ is the upper critical dimension), in two dimensions (middle, $d=2$), and in one dimension (outside, $d=1$). Here, ``ab'' and ``ac'' represent the absorbing and active phases, respectively. For $d \ge d_c$, discontinuous (dashed curve) and continuous transitions (solid line) occur, and they meet at a tricritical point (TP). For $d=2$, a continuous DP transition occurs over the entire region $[0, \kappa_c]$. For $d=1$, a continuous DP transition occurs in the region $[\kappa_*,\kappa_c]$; however, in the interval $[0,\kappa_*]$, the exponent $\alpha$ of the density of active sites $n(t)\sim t^{-\alpha}$ starting from a homogeneous initial state decreases continuously as $\kappa$ is increased for the QDP values.    
\label{fig:fig1}}
\end{figure}

Here, we aim to answer these questions by considering the quantum contact process~\cite{marcuzzi2016, buchhold2017, jo2019, roscher2018, carollo2019, gillman2019, gillman2020, gillman2020_2} in one dimension (1D-QCP) and two dimension (2D-QCP). In the contact process (CP), each element of the system is in an active or inactive state, and its state changes according to the given CP rules ~\cite{marro2005,harris1974,odor2004,ziff1986,dickman1991,henkel2008,hinrichsen2000}. When all the elements are in the inactive state, the system becomes trapped in a frozen configuration, and the dynamics do not proceed further. This absorbing phase transition of the classical CP (CCP) belongs to the directed percolation (DP) universality class. In addition, the DP transition appears in diverse nonequilibrium systems; however, experimental observation has been difficult, although it has been observed in turbulent liquid crystals~\cite{sano2016} and Rydberg atoms~\cite{gutierrez2017}. Here, we focus on the recent observation of the 1D-QCP in a dissipative quantum system of Rydberg atoms. 


The dynamics of the 1D-QCP is described by the Lindblad equation, which consists of a Hamiltonian and dissipative terms. Their contributions to the overall dynamics are adjusted by the model parameters $\omega$ (for the coherent quantum effect) and $\kappa$ (for the incoherent classical dynamics). Thus, the system can exhibit a quantum or classical phase transition in extreme cases. A previous result based on the semiclassical mean-field solution~\cite{marcuzzi2016} showed that the QCP exhibits a continuous (discontinuous) phase transition when $\kappa$ is large (small). Thus, a tricritical point exists, as shown in Fig.~\ref{fig:fig1}. The continuous transition belongs to the mean-field DP (MF-DP) universality class, and the critical behavior at the tricritical point belongs to the mean-field tricritical DP (MF-TDP) class, which occurs at a tricritical point of the tricritical CP (TCP) model. 

Studies using the functional renormalization approach and tensor network approach showed that as the dimension is decreased from the upper critical dimension $d_c=3$~\cite{buchhold2017,jo2019}, the tricritical point shifts toward the quantum axis with $\kappa=0$~\cite{roscher2018,carollo2019}. This result indicates that a quantum phase transition would occur only for $\kappa=0$. This behavior may be related to the fact that quantum phase transitions in equilibrium systems occur only at zero temperature. On the basis of this conjecture, a recent numerical study~\cite{gillman2020} of the 1D-QCP with $\kappa=0$ using the tensor network approach~\cite{carollo2019,gillman2019,gillman2020} revealed that the QCP exhibits a continuous transition in a quantum DP (QDP) class. When the dynamics starts from a homogeneous state, that is, all the sites are in the active state, the density of active sites at time $t$, denoted as $n(t)$, decays as $n(t)\sim t^{-\alpha}$ at a quantum critical point $\omega_c$. The exponent $\alpha_{\rm{QDP}} \approx 0.32$ obtained by the tensor network approach differs from the DP value, $\alpha_{\rm DP}\approx 0.16$. However, the other exponents, except for the hyperscaling exponent $\nu_{\bot}$ of the spatial correlation length, have the DP values. In fact, the discrepancy in $\nu_{\bot}$ is inevitable because the tensor network approach is not valid for strongly entangled quantum systems. Using the quantum jump Monte Carlo (QJMC) approach, we obtain $\nu_{\bot}$ as the DP value, as described later. On the other hand, when the QCP starts from a heterogeneous state, that is, all the sites but one are in the inactive state, all the critical exponents have the classical DP values. Thus, the critical behavior of the 1D-QCP with $\kappa=0$ depends on the initial configuration.


For the 1D-QCP, we are interested in how the quantum fluctuations responds to an classical fluctuations, that is, how the critical exponent $\alpha$ changes to the classical DP behavior as the strength $\kappa$ of the classical fluctuations is increased from zero. We find that there exists an interval $[0,\kappa_*]$ in which the exponent $\alpha$ decreases continuously from the quantum value, $\alpha\approx 0.32$, to the DP value, $\alpha\approx 0.16$, as $\kappa$ is increased from $\kappa=0$ to $\kappa_*$. The phase diagram for the 1D-QCP is shown in Fig.~\ref{fig:fig1}. This result implies that the quantum effect remains to some extent in the region $\kappa=[0,\kappa_*]$. Consequently, the crossover from the 1D-QCP class to the classical DP class occurs in a soft-landing manner. 

We remark that the critical behavior of the TCP model in the mean-field limit, equivalent to the MF-TDP class, exhibits a similar behavior to the 1D-QDP class~\cite{lubeck2006,jo2020}. Only the critical exponent $\alpha$ is also deviated from the DP value and the other exponent values remain the same as the DP values.  One may guess that quantum fluctuations shift the tricritical point onto the quantum transition point $(0,\omega_c)$~\cite{carollo2019}. Thus the first-order transition curve disappears and the exponent $\alpha_{\rm{QDP}}$ has a different value from the corresponding DP value. If this scenario is correct, then the exponent $\alpha$ would have the DP value for any $\kappa > 0$. However, we obtain that there exists an interval in which the soft crossover behavior occurs.

We extend our studies to the 2D-QCP. A previous study~\cite{buchhold2017} showed that the 2D-QCP exhibits a discontinuous transition at $\kappa=0$ using the functional renormalization approach~\cite{buchhold2017}. However, we find that the transition is continuous, and it belongs to the classical DP class. The exponent $\alpha$ for homogeneous initial conditions has the DP value. Thus, the intermediate region $[0,\kappa_*]$ is absent. This suggests that the quantum fluctuations are weaker than the classical fluctuations in two dimensions. For higher dimensions, $d \ge d_c=3$, a discontinuous transition occurs at $\kappa=0$.  

The critical behaviors of the 1D-QCP and 2D-QCP are obtained as follows. The transition point $\omega_c(\kappa)$ for each $\kappa$ value is determined using the neural network (NN) machine learning algorithm. Then, by applying finite-size scaling (FSS) analysis for different system sizes, the exponent $\nu_\bot$ at $\omega_c(\kappa)$ is determined~\cite{carrasquilla2017,kim2018,zhang2019}. At the transition point $\omega_c(\kappa)$, the other critical exponents are determined using the QJMC method~\cite{plenio1998,daley2014} as well as the tensor network method to confirm the results for large systems. 


This paper is organized as follows. First, we introduce the 1D-QCP model and specify the classical and quantum limits in Sec.~\ref{sec:sec2}. The structure of our NN and the optimization scheme are presented in Sec.~\ref{sec:sec3A}. In Sec.~\ref{sec:sec3B}, we present the FSS behaviors obtained using the NN, QJMC, and tensor network methods; the universality class; and the crossover behavior. A summary and final remarks are presented in Sec.~\ref{sec:discussion}. 
Additionally, in the Appendices we present a method of experimental realization of QDP behavior in the Appendix~\ref{appendixA}, the derivation of the model parameter $\kappa$ for the classical dynamics in terms of the experimental parameters of the Rabi frequency $\Omega$ and dephasing rate $\Gamma$ in the Appendix~\ref{appendixB}, the technical details of the QCP in the classical limit using the quantum Monte Carlo method in the Appendix~\ref{appendixC}, the use of the NN approach with different training regions in the Appendix~\ref{appendixD}, the relation $\delta=\alpha$ for the classical CP and classical tricritical CP is presented with other critical exponents in the Appendix~\ref{appendixE}, the continuously varying exponent $\alpha$ is discussed more thoroughly in the Appendix~\ref{appendixF}, and we estimate the critical exponent associated with the temporal correlation $\nu_{\|}$, which is helpful for confirming the critical exponent $\nu_{\bot}=\nu_{\|}/z$ in the Appendix~\ref{appendixG}. 

\section{Model}
\label{sec:sec2}
\begin{figure}
\includegraphics[width=0.8\columnwidth]{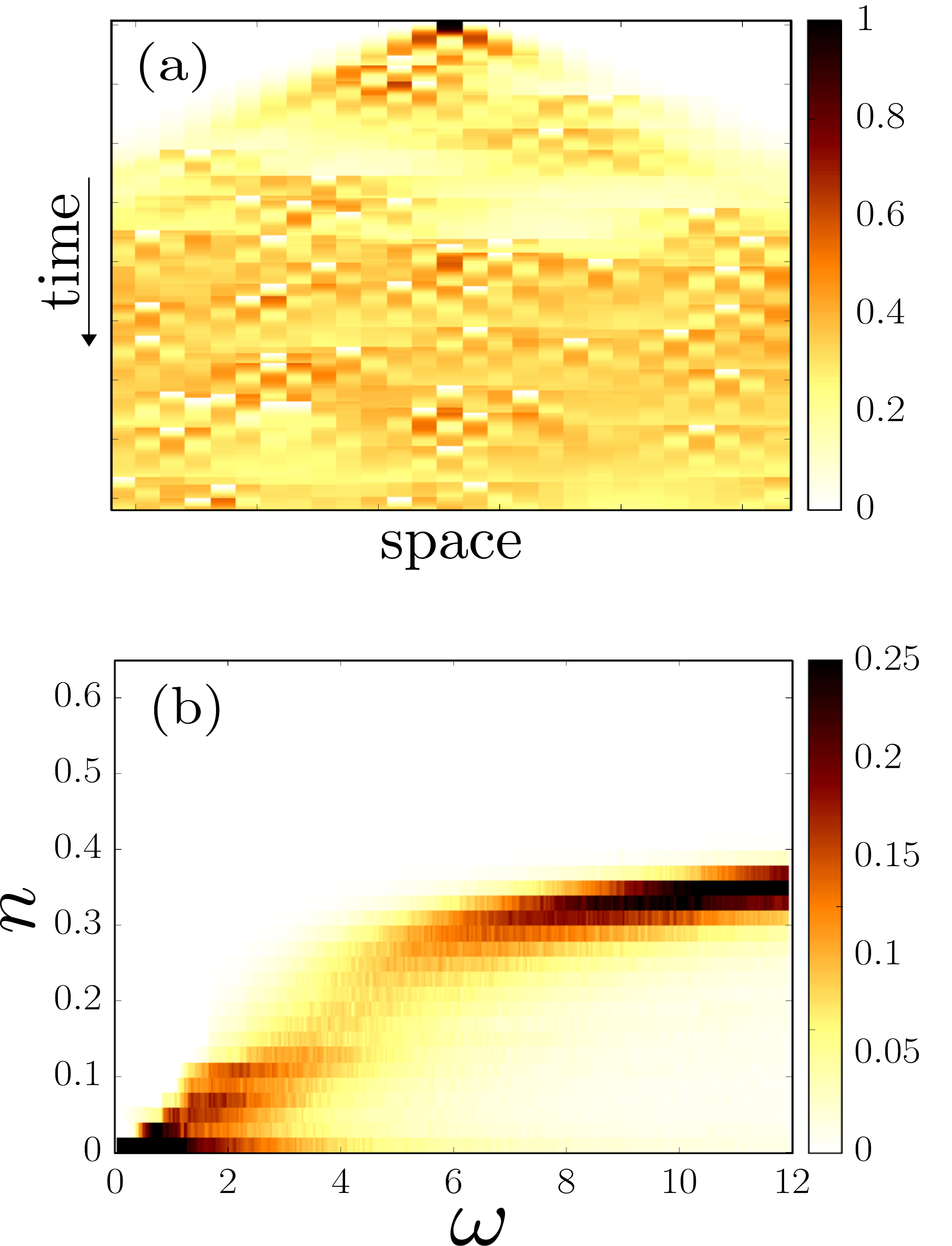}
\caption{Plots obtained using QJMC method. (a) Trajectory of the 1D-QCP for $\kappa= 0$ and $\omega>\omega_c$ from a single active site at the center. (b) Histogram of the densities of active sites in steady states as a function of $\omega$ for system size $N=20$. The data were obtained using QJMC simulations. Time $t$ and control parameter $\omega$ are given in units of $1/\gamma$ and $\gamma$, respectively.}
\label{fig:fig2}
\end{figure}

\begin{figure*}[!t]
\includegraphics[width=2.0\columnwidth]{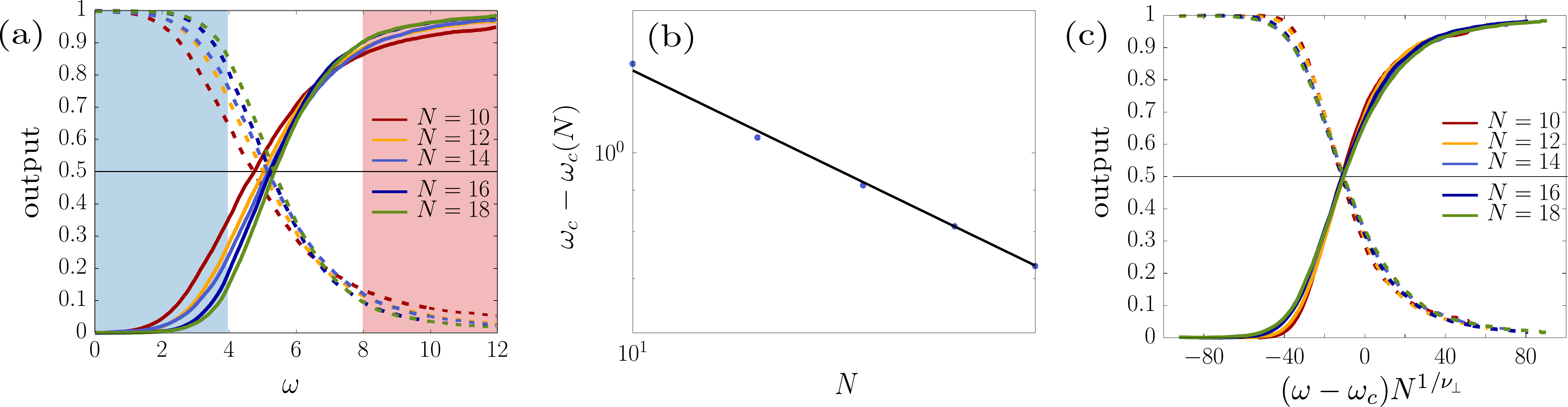}
\caption{Plots obtained using NN method. (a) Plot of the output averaged over a test set as a function of $\omega$ for different system sizes. Solid (dashed) lines represent the values of the first (second) output neuron. From this plot, we estimate the crossing point of the two outputs and regard it as the transition point $\omega_c(N)$ for a given system size $N$. (b) Plot of $\omega_c-\omega_c(N)$ versus $N$, where $\omega_c$ is chosen to yield power-law behavior and is regarded as the transition point in the thermodynamic limit. The slope represents the value of the critical exponent $-1/{\nu_{\bot}}$. (c) Scaling plot of the output versus $(\omega-\omega_c)N^{1/\nu_{\bot}}$. For the obtained numerical values of $\nu_{\bot}$ and $\omega_c$, the data collapse well for system sizes $N = 10, 12, 14, 16$, and $18$. }
\label{fig:fig3}
\end{figure*}

We consider a one-dimensional quantum spin chain with a periodic boundary condition, where the active and inactive states of a site represent the up and down spin states, denoted as $\left| \uparrow \rangle\right.$ and $\left|\downarrow\rangle\right.$, respectively. A QCP consists of three incoherent and two coherent processes: i) decay, in which an active site is incoherently inactivated spontaneously at a rate $\gamma$; ii) incoherent branching or coagulation, in which an active particle incoherently activates or inactivates an inactive particle at the nearest-neighbor site at a rate $\kappa$, respectively; and iii) coherent branching or coagulation, which is a quantum counterpart of process ii) driven by a Hamiltonian at a rate $\omega$. The classical ii) and quantum iii) rules are in competition, which may change the transition behavior. 

The time evolution of the density matrix $\hat{\rho}$ is described by the Lindblad equation, which consists of the Hamiltonian and dissipative terms~\cite{breuer2002}:
\begin{align}
\label{eq:lindqcp}
\partial_t\hat{\rho}&=-{\rm i}\left[ \hat{H}_S,\hat{\rho} \right]
+ \sum_{a=d,b,c}\sum_{\ell=1}^N\left[ \hat{L}^{(a)}_{\ell}\hat{\rho} \hat{L}^{(a)\dagger}_{\ell}
-\frac{1}{2} \left\{ \hat{L}^{(a)\dagger}_{\ell}\hat{L}^{(a)}_{\ell},\hat{\rho} \right\} \right]\,.
\end{align}
The Hamiltonian $\hat{H}_S$, which governs the branching and coagulation processes and represents coherent interactions, is expressed as
\begin{align}
\label{eq:QCP_hamiltonian}
\hat{H}_S= \omega \sum_{\ell=1}^{N}\Big[ (\hat{n}_{\ell-1}+\hat{n}_{\ell+1})\, \hat{\sigma}^x_{\ell} \Big]\,.
\end{align}
Here, $\hat{\sigma}^i_l$ denotes the Pauli matrix, where the superscript and subscript represent the spin axis and site index, respectively, and $\hat{n}_l$ is the number operator for an up spin at the $l$th site.
The Lindblad decay, branching, and coagulation operators are given by 
\begin{align}
\label{eq:QCP_decay}
\hat{L}_{\ell}^{(d)}&=\sqrt{\gamma}\,\hat{\sigma}_{\ell}^{-}\,, \\
\label{eq:QCP_branching}
\hat{L}_{\ell}^{(b)}&=\sqrt{\kappa}\,(\hat{n}_{\ell-1}+\hat{n}_{\ell+1})\hat{\sigma}^+_{\ell}\,,\\ 
\label{eq:QCP_coagu}
\hat{L}_{\ell}^{(c)}&=\sqrt{\kappa}\,(\hat{n}_{\ell-1}+\hat{n}_{\ell+1})\hat{\sigma}^-_{\ell}\,,
\end{align}
respectively. 
$\hat{\sigma}^{+}_{\ell}$ and $\hat{\sigma}^{-}_{\ell}$ are the raising and lowering operators of the spin at site $\ell$, which are defined in terms of the spin basis as $\hat{\sigma}^{+} = \left| \uparrow \rangle \langle \downarrow \right|$ and
$\hat{\sigma}^{-} = \left| \downarrow\rangle \langle \uparrow \right|$, respectively. 
In addition, $\hat{n}=\hat{\sigma}^{+}\hat{\sigma}^{-}$ and $\hat{\sigma}^x=\hat{\sigma}^{+}+\hat{\sigma}^{-}$ are the number operator and spin flip operator, respectively. We rescale the time and quantum control parameters $\omega$ and $\kappa$, respectively, in units of $\gamma$; therefore, we set $\gamma=1$.

Quantum branching and coagulation occur at a rate $\omega$, and the corresponding classical processes occur at a rate $\kappa$. When $\omega\to 0$, the model is reduced to the CCP, which belongs to the DP class. Here, we first consider the pure quantum limit $\kappa\to 0$, but with finite $\omega$. The opposite limit, $\omega\to 0$ with finite $\kappa$, and the case of finite $\omega$ and $\kappa$ are discussed in the Appendix~\ref{appendixC}. 

When $\omega$ is small, inactive particles become more abundant with time, and eventually the system is fully occupied by inactive particles. Thus, the system is no longer dynamic and falls into an absorbing state, which is represented by $\hat{\rho}_{\text{ab}}= \left| \downarrow\cdots\downarrow\rangle \langle \downarrow\cdots\downarrow\right|$. When $\omega$ is large, the system remains in an active state with a finite density of active particles [Fig.~\ref{fig:fig2}(a)]. Thus, the QCP exhibits a phase transition from an active to an absorbing state as $\omega$ is decreased.

\section{RESULTS}

\subsection{NN approach}
\label{sec:sec3A}

\begin{figure}[!t]
	\centering
	\includegraphics[width=0.9\linewidth]{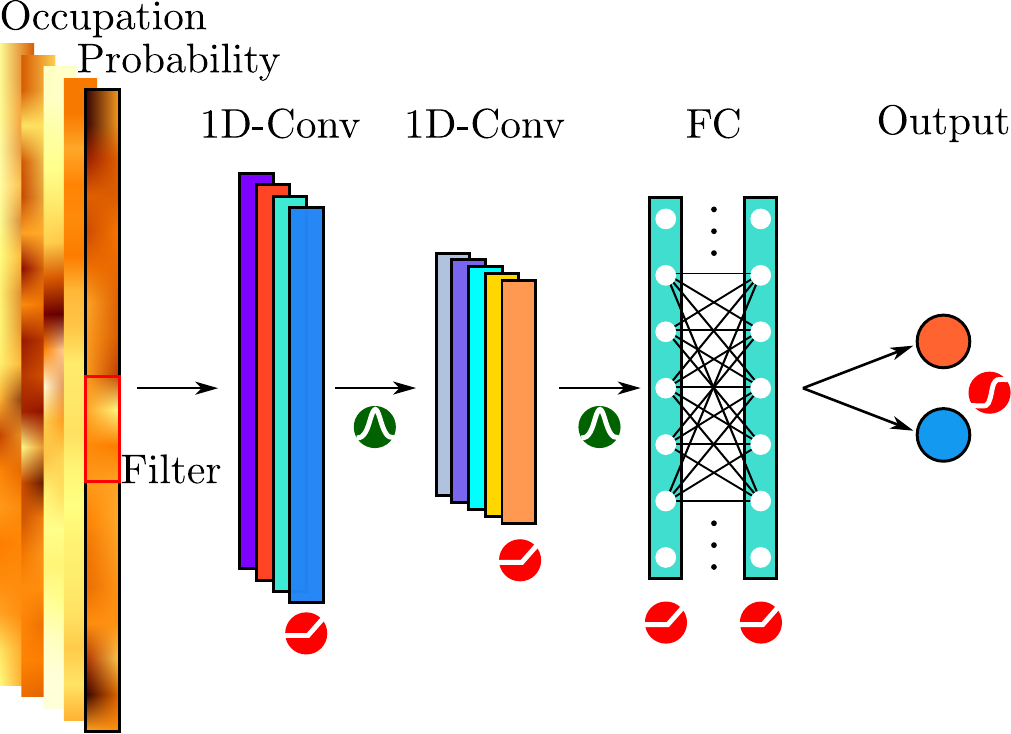}
	\caption{
		Schematic illustration of the convolutional NN built in combination with a one-dimensional convolutional layer (1D-Conv) and a fully connected layer (FC). 
		The red circles represent the activation function of each layer.
		The green circles below the arrows represent the batch normalization. 
	}
	\label{fig:fig4}
\end{figure}





The NN approach has recently been used as a powerful tool~\cite{lecun2015, carleo2019} for understanding phase transitions in classical systems~\cite{carrasquilla2017}, which exhibit patterns involving many components. Each component has one of two values, for instance, the up and down spin states in ferromagnetic systems. 
By contrast, each component of a quantum system has a real value, and thus the patterns are much more complex. Nevertheless, the NN approach has reportedly been used successfully to determine the transition points of closed quantum systems on the basis of simulation data~\cite{broecker2017,venderley2018,canabarro2019} and experimental images~\cite{rem2019,bohrdt2019}. 
In addition, an unsupervised NN approach was used to generate the configurations of quantum dissipative systems in the steady state ~\cite{schuld2019,yoshioka2019,hartmann2019,nagy2019,vicentini2019}. 
This approach was efficient, because it uses less computing resources to generate configurations compared to conventional simulations.
However, determining the transition point accurately in dissipative quantum systems requires large system sizes, and thus the simulations require considerable computational resources. Here, we use the NN approach as an alternative method and obtain a transition point that is sufficiently accurate for investigation of the critical behavior of the 1D-QCP at the transition point.

For classical systems, the transition point of a continuous absorbing transition is usually indicated by the presence of power-law behavior of the order parameter with respect to time~\cite{marro2005,jo2020}. Consequently, a large system size is required to identify the transition point. Accurately identifying the transition point using QJMC simulations of the QCP is even more difficult and is thus a challenging problem. To overcome this difficulty, we note that the system is in the absorbing state for $\omega\ll \omega_{c}$ and in the active state for $\omega \gg \omega_{c}$. Combining this observation with a recently proposed NN supervised learning concept, we identify the transition point as follows.

To implement the NN approach, we first organize a dataset of the occupation probability of site $\ell$, which is denoted as $p_\ell(t)=\text{Tr} [\hat{\rho}(t)\hat{n}_\ell]$. Using the QJMC method, we generate a steady-state configuration and obtain the occupation probabilities of each site, $\{p_\ell\}$. We collect $5\times 10^3$ configurations in $\omega\in [0, 12]$ at $\Delta\omega=0.04$ intervals. To prepare the training dataset for supervised learning, we label the configurations using one-hot encoding~\cite{harris2012}, where the absorbing state ($\omega\in [0,4]$) is encoded as $(0,1)$, and the active state ($\omega\in [8,12]$) is encoded as $(1,0)$ [see shaded regions in Fig.~\ref{fig:fig3}(a)]. 

After we collect the snapshots, we try to train the NN.
The objective of the learning procedure is to optimize the NN to adjust the weights of the connections between neural units to achieve the variational minimization of a properly defined cost function.
To this end, we construct the hidden layers of the NN, including one-dimensional convolutional layers, batch normalization layers~\cite{ioffe2015}, and fully connected layers, as shown in Fig.~\ref{fig:fig4}. We employ the framework of \textsc{tensorflow}~\cite{abadi2016} and use ReLU and $\tanh$ for the activation function in the hidden layer. Two neurons in the output layer are used, and a softmax function is used as the activation function in the output layer. We employ the cross-entropy or the mean-square error function as the cost (error) function of the NN, which is then optimized using Adam~\cite{kingma2014} or RMSProp. We vary the architecture and optimization algorithms in various ways. Regardless of these changes, the well-trained machines produce consistent results. To check the sensitivity to the positions of the left and right boundaries, the NN approach with different training regions is described, and its results are presented in the Appendix~\ref{appendixD}.
When the NN is well-trained with the labeled training dataset in the two regions, we obtain the outputs for the entire $\omega$ region. 

Next, using the obtained transition points $\omega_c(N)$ for selected system sizes, we perform FSS analysis and identify the transition point in the thermodynamic limit, $\omega_c$. We also determine the correlation length exponent $\nu_\bot$. Next, we determine the other critical exponents by performing extensive QJMC simulations up to system size $N=20$ and applying the tensor network method to a large system ($N=80$) at $\omega_c$. Specifically, the tensor network method is implemented by the QJMC method with the matrix product states and time-evolving block decimation.

\subsection{FSS analysis of 1D-QCP}
\label{sec:sec3B}

In Fig.~\ref{fig:fig3}(a), the two outputs of the NN indicate the probabilities that the system will fall into the absorbing state and remain in the active state, respectively. The crossing point of these outputs indicates a transition point $\omega_c(N)$ for a given system size $N$ [Fig.~\ref{fig:fig3}(a)]. Several studies~\cite{carrasquilla2017,kim2018,zhang2019} have shown that the predictability exhibits FSS behavior. Using the obtained $\omega_c(N)$ for different system sizes, we determine $\omega_c$ in the thermodynamic limit by plotting $\omega_c-\omega_c(N)$ versus $N$ [Fig.~\ref{fig:fig3}(b)], which is expected to behave as $\omega_c-\omega_c(N)\sim N^{-1/\nu_{\bot}}$. Indeed, the plot exhibits power-law behavior when an appropriate value of $\omega_c$ is chosen, and the critical exponent $\nu_{\bot}$ is obtained from the slope of the power-law curve. We obtain $\omega_c \approx 6.04$ and $\nu_{\bot}=1.06\pm 0.04$; the latter is consistent with the value of $\nu_{\bot}\approx 1.096$ for the DP class in one dimension but differs from the value of $\nu_{\bot}\approx 0.5\pm 0.2$ obtained using the tensor network approach. The scaling plot is drawn in the form of the output versus $(\omega-\omega_c)N^{1/\nu_{\bot}}$ for different $N$ [Fig.~\ref{fig:fig3}(c)]. The data for different system sizes seem to collapse. 

\begin{figure}[t!]
\includegraphics[width=1.00\columnwidth]{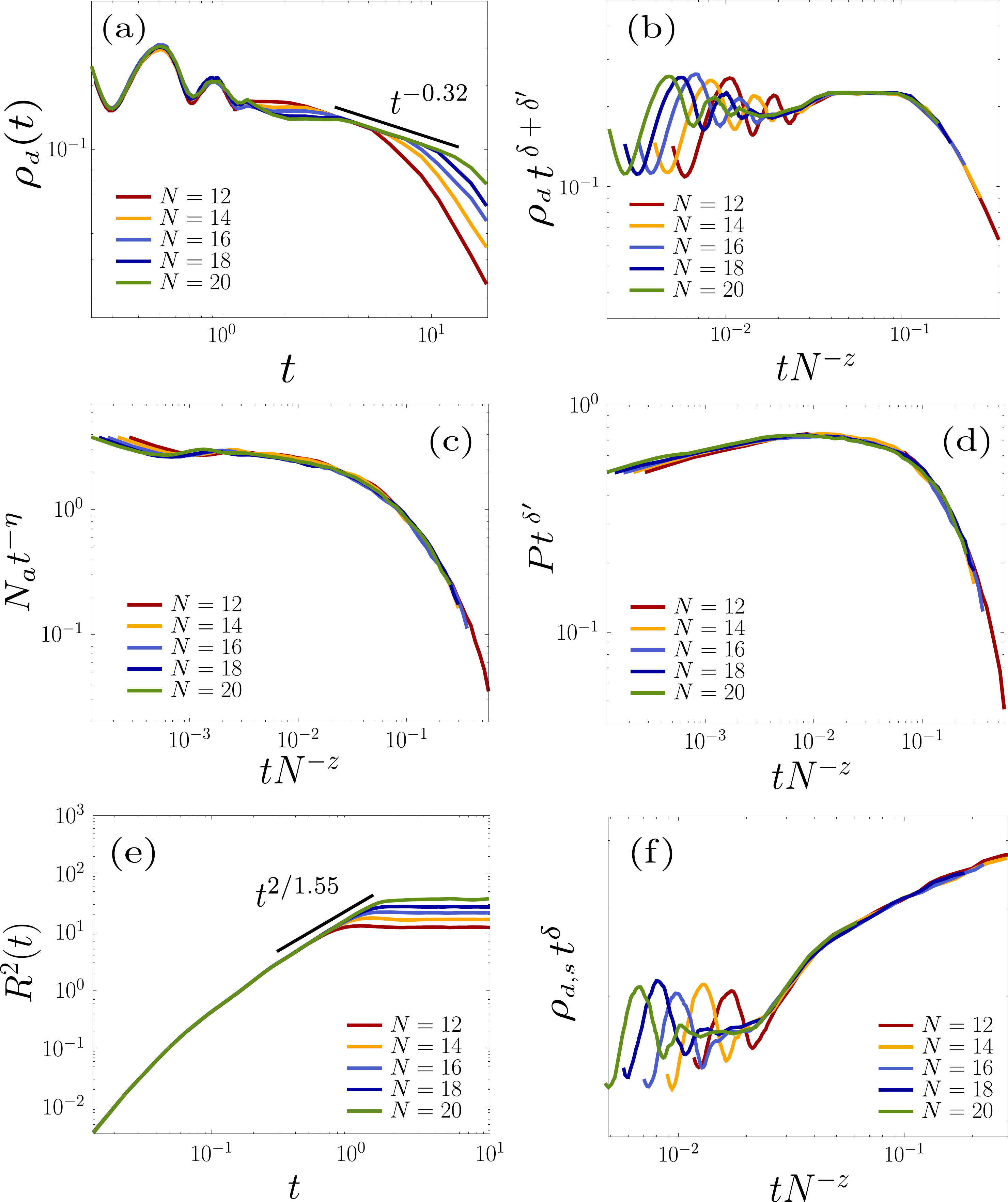}
\caption{Estimates of the critical exponents of the 1D-QCP starting from a single active site. (a) Plot of $\rho_d(t)$ versus $t$, which behaves as $\rho_d(t)\sim t^{-\delta-\delta'}$. (b) Scaling plot of $\rho_d(t)t^{\delta+\delta'}$ versus $tN^{-z}$ for $\delta+\delta'=0.32$ and $z=1.55$.
(c) Scaling plot of $N_a(t) t^{-\eta}$ versus $tN^{-z}$ for $\eta=0.30$ and $z=1.55$.
(d) Scaling plot of $P(t)t^{\delta'}$ versus $tN^{-z}$ for $\delta' =0.16$ and $z=1.55$.
(e) Plot of $R^2(t)$ as a function of $t$. 
(f) Scaling plot of $\rho_{d,s}(t)t^{\delta}$ versus $tN^{-z}$ for $\delta=0.16$ and $z=1.55$.}
\label{fig:fig5}
\end{figure}

Next, we measure the values of the other critical exponents using the QJMC method in the critical region around $\omega_c$ with the tensor network method for larger system sizes. First, we consider an initial state in which a single active seed is present at $\ell=0$, and the remaining sites are inactive. This configuration is expressed as $\hat{\rho}(0)=\hat{\sigma}^+_0\hat{\rho}_{\text{ab}}\hat{\sigma}^-_0$. We measure the following quantities: 
i) the survival probability, that is, the probability that the system does not fall into an absorbing state, $P(t)=1-\text{Tr}[\hat{\rho}(t)\hat{\rho}_{\text{ab}}]$; ii) the number of active sites, $N_a(t)=\sum_{\ell}\text{Tr}[\hat{\rho}(t)\hat{n}_{\ell}]$; iii) the mean square distance of the active sites from the origin, $R^2(t)=\sum_{\ell}\text{Tr}[{\ell}^2\,\hat{\rho}(t)\hat{n}_{\ell}]/N_a(t)$;
iv) the seed-site density over all runs, $\rho_d(t)=\text{Tr}[\hat{\rho}(t)\hat{n}_{\ell=0}]\sim N_a(t)/R(t)$; and v) the seed-site density over surviving runs, $\rho_{d,s}(t)=\rho_d/P(t)$. 
At the transition point, these quantities exhibit the following power-law behaviors: 
$P(t)\propto t^{-\delta^\prime}$, $N_a(t)\propto t^{\eta}$, $R^2(t)\propto t^{2/z}$, $\rho_{d}(t)\propto t^{\eta-1/z}$, and $\rho_{d,s}(t)\propto t^{-\delta}$. 
For the relation $\rho_d(t)= \rho_{d,s}(t) P(t)\sim t^{-\delta-\delta^\prime}$, the scaling relation $\eta-1/z=-(\delta+\delta')$ holds. We estimate the exponents $\delta+\delta'$, $\eta$, $\delta'$, $z$, and $\delta$ by direct measurement of the slopes in the double-logarithmic plots, as shown in Fig.~\ref{fig:fig5}. We estimate the exponent $z$ using the data collapse technique. For instance, for the survival probability $P(t)$, we plot $P(t)t^{\delta^\prime}$ versus $tN^{-z}$ for different system sizes $N$. We determine $z$ as the value at which the data for different system sizes collapse onto a single curve.   
The values of the critical exponents are in good agreement with the DP values within the error bars (Table I).

\begin{table}[h!]
\begin{center}
\caption{Critical point and critical exponents for the 1D-QCP.
}
\setlength{\tabcolsep}{8pt}
{\renewcommand{\arraystretch}{1.0}
\begin{tabular}{ccccc}
    \hline
    \hline
	& 1D-QCP from & 1D-QCP from  & \multirow{2}{*}{1D-DP} \\
	&  CNN\footnote{Convolutional NN}+QJMC  &  tensor network~\cite{gillman2019, carollo2019} &  \\
   \hline
    $\omega_c$ & $6.04$& $6.0\pm 0.05$ & ---\\
    $\delta'$ & $0.16\pm 0.05$& $0.26\pm 0.04$& $ 0.159$\\
    $z$ & $1.55\pm 0.06$ & $1.61\pm 0.16$& $1.581$\\
    $\eta$   & $0.30\pm 0.05$ & $0.26\pm 0.05$& $0.313$\\
    $\delta+\delta'$   & $0.32\pm 0.01$& $0.36\pm 0.12$ & $0.318$\\
    $\alpha$   & $0.32\pm 0.01$& $0.36\pm 0.08$ & $0.159$\\
    $\nu_{\bot}$  & $1.06\pm 0.04 $& $0.5\pm 0.2$ & $1.096$\\
    \hline
    \hline
\end{tabular}}
\label{tab:tab1}
\end{center}
\end{table}

\begin{figure*}[!t]
\includegraphics[width=2.00\columnwidth]{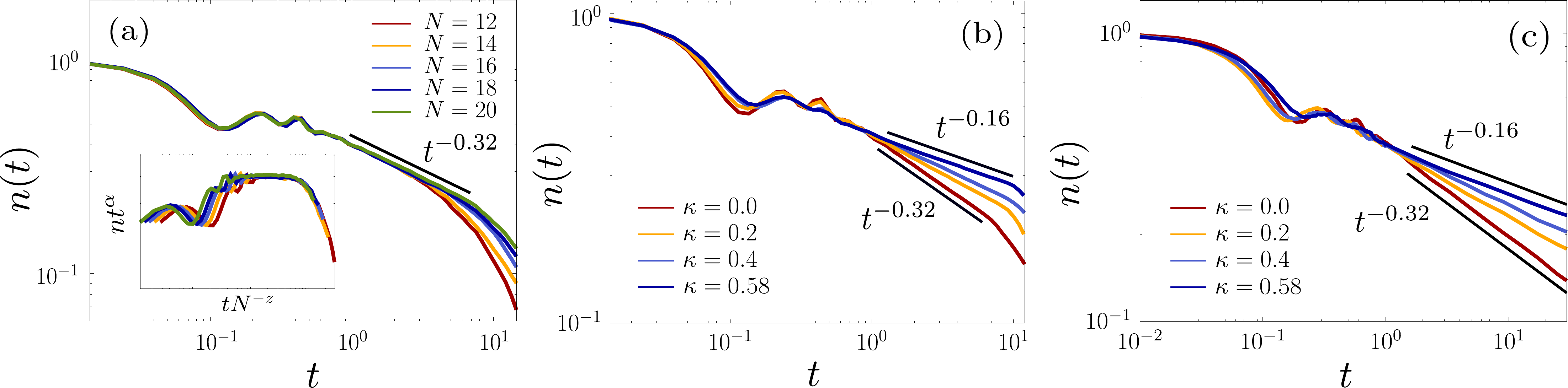}
\caption{Estimates of the critical exponents $\alpha$ starting from the homogeneous state. (a) Plot of $n(t)$ as a function of $t$ for different system sizes when $\kappa=0$, which shows that $n(t)\sim t^{-\alpha}$, with $\alpha=0.32$. 
Plot of $n(t)$ as a function of $t$ for different $\kappa$ in the range $\kappa\in[0, 0.6]$ in steps of $0.2$ using (b) QJMC with $N=20$ and (c) tensor network method with $N=80$ and bond dimension $\chi=1024$. The lower (upper) solid line is a guideline with slope $-0.32$ ($-0.16$).
\label{fig:fig6}}
\end{figure*}

Second, we take a homogeneous initial state in which the entire system is occupied by active sites at $t=0$, which is expressed as $\hat{\rho}(0)= \left| \uparrow\cdots\uparrow\rangle \langle \uparrow\cdots\uparrow\right|$. From this initial state, we measure vi) the density $n(t)$ of active sites at time $t$ averaged over all runs. This quantity is formulated as $n(t)=(\sum_{\ell}\text{Tr}[\hat{\rho}(t)\hat{n}_{\ell}])/N$. We find that $n(t)$ exhibits power-law decay as $n(t)\sim t^{-\alpha}$ with exponent $\alpha=0.32\pm 0.01$, as shown in Fig.~\ref{fig:fig6}(a). This value is consistent with the result obtained by applying the tensor network approach; however, it is not consistent with the corresponding DP value, which was estimated as $\alpha_{\rm DP}=0.16$. Therefore, the 1D-QCP for $\kappa=0$ creates a new type of universal behavior. 


We note that $\rho_d(t)$ and $n(t)$ are actually the same quantity even though they emerge from different initial states~\cite{mendes1994}. They exhibit the same critical behaviors in the CP class (See the Appendix~\ref{appendixE}), but they exhibit different critical behaviors for the 1D-QCP. This behavior is unusual because the universality class is independent of the initial state according to the theory of critical phenomena. To understand the underlying mechanism, we increase the control parameter $\kappa$ from zero to 0.6 in steps of $0.2$ and explore the behavior of $n(t)$ at each $\omega_c(\kappa)$~[Fig.~\ref{fig:fig6}(b)]. We find that the value of $\alpha$ decreases continuously from $0.32$ for $\kappa=0$ to $\alpha=0.16$ for $\kappa=0.6$. 
Furthermore, we apply the tensor network method based on the matrix product states and time-evolving block decimation to confirm the results for a large system~[Fig.~\ref{fig:fig6}(c)]. Using FSS analysis, we determine the exponent $\alpha$ for each $\kappa$ (See the Appendix~\ref{appendixF}). These results suggest that $\alpha$ decreases continuously as $\kappa$ is increased and reaches the DP value at $\kappa_*\approx 0.58$ (Table~\ref{tab:tab2}). 

\subsection{FSS analysis of 2D-QCP}
\label{sec:sec3B} 
\begin{figure}[!t]
\includegraphics[width=1.0\columnwidth]{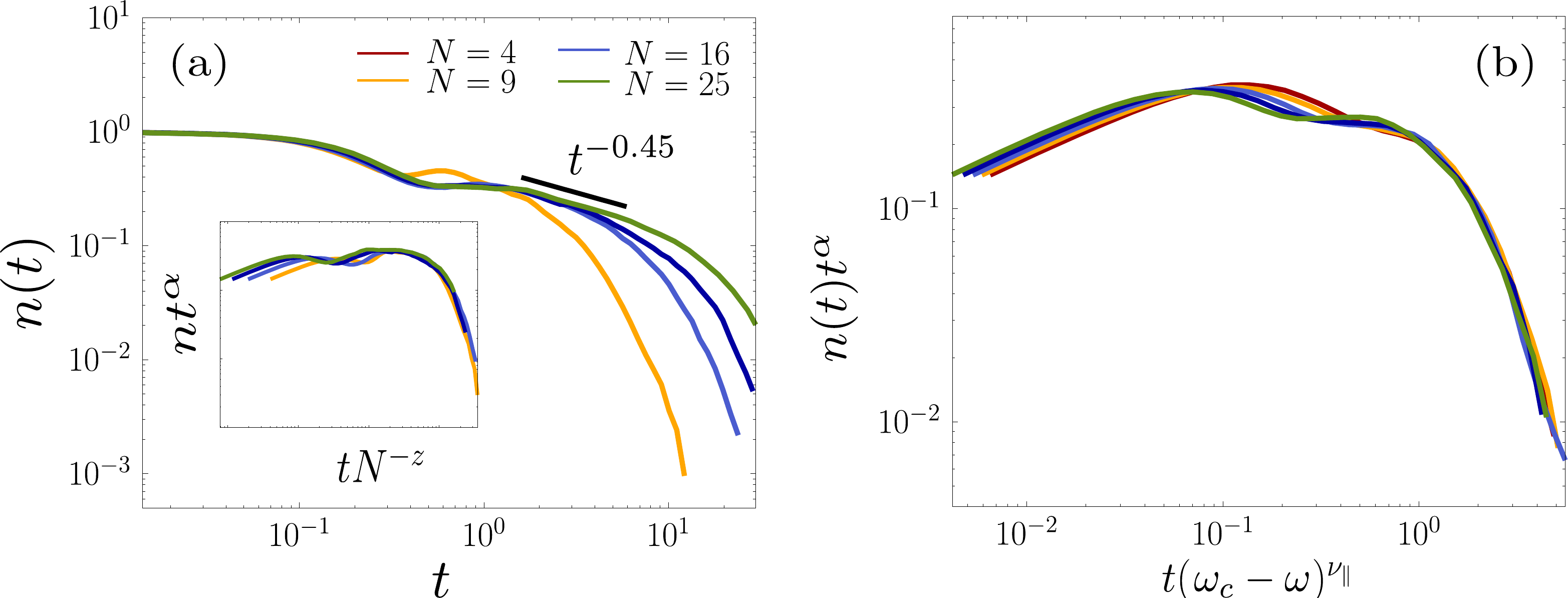}
\caption{Estimates of the critical exponents $\alpha$ and $\nu_{\|}$ for the 2D-QCP. Initial state is homogeneous. (a) Plot of $n(t)$ as a function of $t$ for different system sizes when $\kappa=0$, which shows that $n(t)\sim t^{-\alpha}$, with $\alpha=0.45$. Inset: scaling plot of $n(t)t^{\alpha}$ versus $tN^{-z}$ for $\alpha=0.32$ and $z=1.76$.
(b) Data points collapse well onto a single curve for $\omega_c=0.94$, $\alpha=0.45$, and $\nu_{\|}=1.30$. The units of the control parameter are given as $\gamma$.
\label{fig:fig7}}
\end{figure}

We investigate the critical behavior of the 2D-QCP. At $\kappa=0$, we find that the 2D-QCP exhibits a continuous absorbing-state phase transition. By a process similar to that used for the 1D-QCP, we obtain the critical exponents of the 2D-QCP using QJMC simulations. For $\kappa=0$, we obtain $\omega_c\approx 0.94$. At this $\omega_c$, we find that $n(t)$ exhibits power-law decay as $n(t)\sim t^{-\alpha}$ with $\alpha=0.45\pm 0.03$, as shown in Fig.~\ref{fig:fig7}(a). This value is in agreement with the corresponding DP value in two dimensions. In Fig.~\ref{fig:fig7}(b), the exponent $\nu_{\|}$ is determined by rescaling the plot of $n(t)t^{\alpha}$ versus $t(\omega_c-\omega)^{\nu_{\|}}$ for different $\omega$ as $\nu_{\|}=1.30$. Thus, $\nu_{\bot}=\nu_{\|}/z\simeq 0.74$, and the critical exponents obtained for a homogeneous initial state are in good agreement with the DP values within the error bars.

\begin{figure}[!t]
\includegraphics[width=1.0\linewidth]{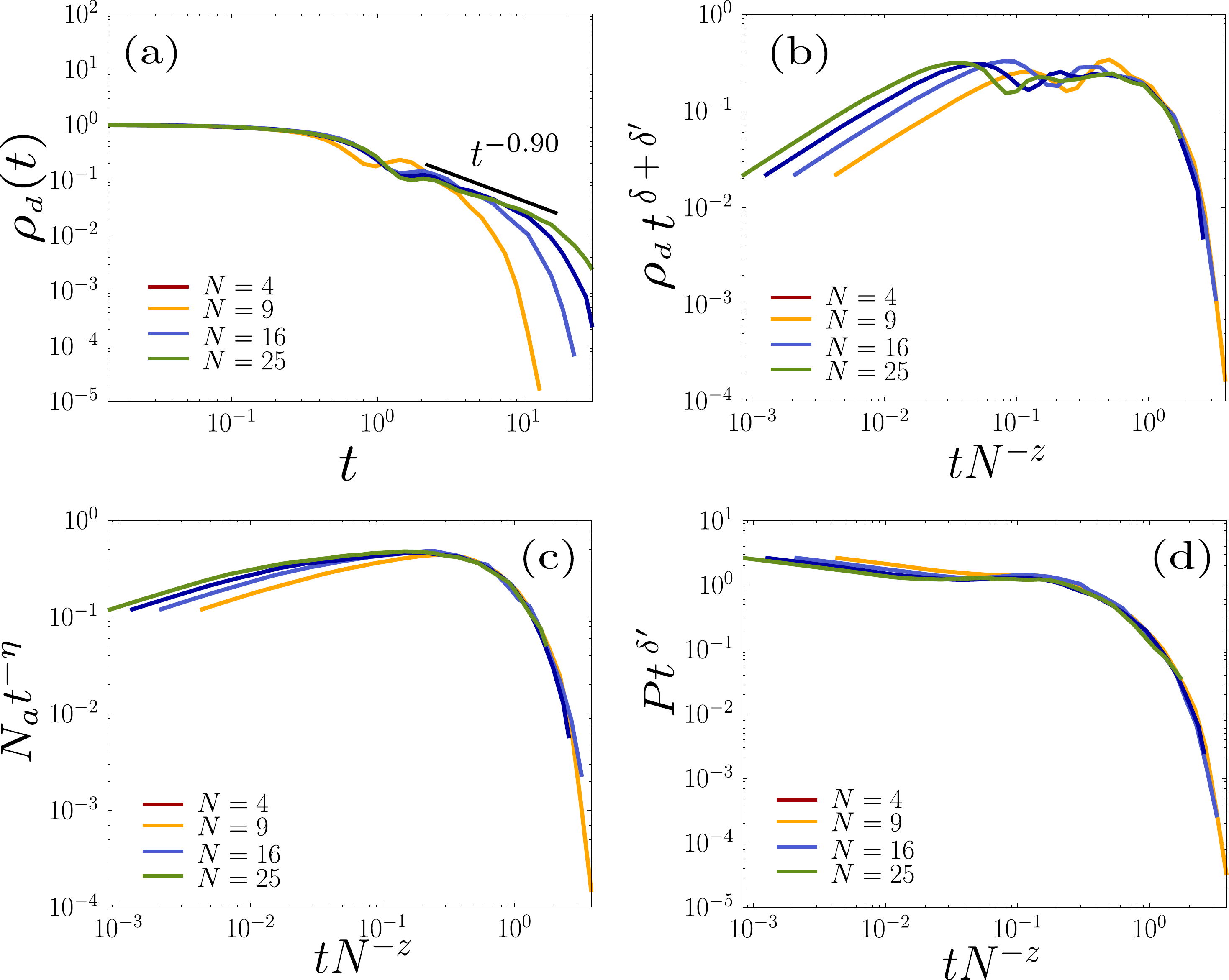}
\caption{
Estimates of the critical exponents of the 2D-QCP starting from a single active site. 
(a) Plot of $\rho_d(t)$ versus $t$, which behaves as $\rho_d(t)\sim t^{-\delta-\delta'}$. (b) Scaling plot of $\rho_d(t)t^{\delta+\delta'}$ versus $tN^{-z}$ for $\delta+\delta'=0.90$ and $z=1.76$.
(c) Scaling plot of $N_a(t) t^{-\eta}$ versus $tN^{-z}$ for $\eta=0.23$ and $z=1.76$.
(d) Scaling plot of $P(t)t^{\delta'}$ versus $tN^{-z}$ for $\delta'=0.45$ and $z=1.76$.}
\label{fig:fig8}
\end{figure}

Next, we estimate the exponents $\eta$ and $\delta'$ by direct measurement of the slopes of the double-logarithmic plots, as shown in Fig.~\ref{fig:fig8}. We estimate the exponent $z$ using the data collapse technique. For instance, for the survival probability $P(t)$, we plot $P(t)t^{\delta^\prime}$ versus $tN^{-z}$ for different system sizes $N$. We determine $z$ as the value at which the data for different system sizes collapse onto a single curve. For the 2D-QCP, we find that rapidity-reversal symmetry holds because $\alpha=\delta'$.
The values of the critical exponents are in good agreement with the DP values within the error bars. Therefore, the 2D-QCP for $\kappa=0$ belongs to the DP class. We conclude that the quantum coherent effect is irrelevant in two dimensions.

\begin{table}[!h]
\begin{center}
\caption{Critical exponent $\alpha$ for different $\kappa$ values.
}
\vskip 2mm
\setlength{\tabcolsep}{25pt}
{\renewcommand{\arraystretch}{1.3}
\begin{tabular}{cc}
    \hline
    \hline
	 $\kappa$ & $\alpha$ \\
   \hline
    $0.0$ & $0.32\pm 0.01$  \\
    $0.1$ & $0.28\pm 0.01$ \\
    $0.2$ & $0.24\pm 0.01$ \\
    $0.3$ & $0.22\pm 0.01$ \\
    $0.4$ & $0.20\pm 0.01$ \\
    $0.5$ & $0.18\pm 0.01$ \\ \hline
    $\ge 0.58$ &  {1D-DP values}\\
    \hline
    \hline
\end{tabular}}
\label{tab:tab2}
\end{center}
\end{table}

\section{summary and DISCUSSION}
\label{sec:discussion}

We investigated the 1D-QCP and 2D-QCP as prototypical examples of nonequilibrium absorbing phase transitions in dissipative quantum systems. A phase diagram was obtained (Fig.~\ref{fig:fig1}) in the parameter space ($\kappa$, $\omega$), where these variables represent the contributions of the classical and quantum effects, respectively. When the 1D-QCP starts from a homogeneous state, the transition curve between the absorbing and active phases has two parts: a quantum region $[0,\kappa_*]$ and the classical DP region $[\kappa_*, \kappa_c]$. In the quantum region, the critical exponent $\alpha$, which is associated with the density of active sites $n(t)$, decreases continuously as $\kappa$ is increased from the quantum value to the classical DP value, as presented in Table II. Thus, the crossover from quantum to classical behavior proceeds gradually. When the 1D-QCP starts from a heterogeneous state, this soft-landing crossover does not occur. For the 2D-QCP, we find a continuous transition at $\kappa=0$ in the DP class, which is inconsistent with the prediction by the functional renormalization group approach~\cite{buchhold2017}. Interestingly, in the mean-field solution~\cite{buchhold2017, marcuzzi2016, jo2019}, the transition in the region near $\kappa=0$ is discontinuous. This discontinuous transition changes to a continuous transition at a tricritical point, which belongs to the MF-TDP class.  In the MF-TDP class, the critical exponent $\alpha_{\rm MF-TDP}$ has a different value from the corresponding DP value. However, there is no crossover region. Thus, there is a discrepancy in the crossover behavior between the 1D-QDP and the mean-field QDP. It would be interesting to investigate underlying mechanism for this discrepancy, which is left for the next problem.

The NN approach we used was applied to a dataset obtained by QJMC simulations. 
We believe that this approach can be applied directly to snapshots obtained in cold Rydberg atom experiments. Furthermore, our approach will be used in the near future for other dissipative systems where a transition point is hardly determined. Examples include the dissipative transverse-field Ising model, dissipative XYZ model, and dissipative anisotropic Heisenberg model.

\begin{acknowledgments}
This research was supported by the NRF, Grant No.~NRF-2014R1A3A2069005 and the AI Institute of Seoul National University (AIIS) through its AI Frontier Research Grant in 2020 (BK). MJ would like to thank E. Gillman for helpful discussions. 
\end{acknowledgments}

\begin{appendix}

\section{Experimental realization}

\label{appendixA}

Here, we concern experimental realization of QCP, specifically on the antiblockade mechanism matching to the branching and coagulation processes. It is known theoretically that the quantum limit $\kappa=0$ can be reached by controlling the dephasing strength. 
This is realized by controlling the laser linewidths and residual Doppler broadening. The classical DP limit (strong dephasing limit) of QCP is experimentally verified in Ref.~\cite{gutierrez2017}; however, it is intrinsically unreachable to the quantum limit with zero dephasing, because setting the laser linewidth to zero is challenging. Thus, achieving the experimental realization of the quantum limit has yet to be resolved. It is analogous to the situation that zero temperature is unreachable precisely experimentally in the transverse Ising model. Here, we propose that using our result, a quantum effect can be found at $\kappa \ne 0$ but in the interval $[0, \kappa_*]$. Below we derive the formula to reach this interval in terms of experimental quantities.

An essential factor in this experiment is the antiblockade effect, in which an inactive (active) spin is activated (inactivated) by detuning the excitation energy so that it is comparable to the interaction energy with the active spin of the nearest neighbor. Moreover, when the dephasing noise is sufficiently strong, the quantum coherences are suppressed, and the dynamics may be reduced to those of the CCP process\mbox{~\cite{marcuzzi2015}}. Otherwise, quantum coherence is effective, and a combination of coherent and incoherent CPs can be realized\mbox{~\cite{marcuzzi2016,buchhold2017}}. Thus, by controlling the ratio of the Rabi frequency and the dephasing rate, one may observe crossover behavior from a quantum to a classical nonequilibrium phase transition. It has been shown that the classical parameter is related to $\kappa=4\Omega^2/\Gamma$, where $\Omega$ is the Rabi frequency, and $\Gamma$ is the dephasing rate\mbox{~\cite{marcuzzi2015}}. Thus, for $\Omega <\sqrt{\kappa_*\Gamma}/2$, QDP behavior may appear. This region is a quantum critical region where quantum fluctuations play a role in the universal behavior.

Let us consider the Rydberg atom represented by two spin states, $\mid\downarrow\rangle$ and $\mid\uparrow\rangle$, which are the eigenstates of the Pauli matrix in the $z$ direction, where the up spin state indicates Rydberg excitation. We start with the Hamiltonian in the rotating-wave approximation of $N$ coupled Rydberg atoms on a lattice, as follows:
\begin{equation}
\hat{H}_{R}=\Omega\sum_\ell^N \hat{\sigma}_\ell^x+
\Delta\sum_\ell^N \hat{n}_\ell+\sum_{\ell\neq m}\frac{V_{\ell m}}{2}\hat{n}_\ell \hat{n}_m,
\label{eq:HR}
\end{equation}
where $\Omega$ and $\Delta$ are the Rabi frequency induced by the external laser
and the detuning strength, respectively. 
The third term on the r.h.s. of Eq.~(\ref{eq:HR}) describes the interaction between the up spins, where
$V_{\ell m}$ is a power-law decay function of distance: 
\begin{align}
V_{\ell m}=\frac{C_p}{|{\bf x}_l-{\bf x}_m|^p},
\label{eq:int}
\end{align}
where $C_p$ is the dispersion coefficient~\cite{saffman2010}, and 
${\bf x}_l$ is the position of the $l$th site.
Note that $p$ characterizes the interaction~\cite{saffman2010}; 
$p=3$ for the dipole interaction associated with
$d$-orbital excitation, and $p=6$ for the van der Waals interaction with $s$-orbital excitation. For $p=6$ in Eq.~(\ref{eq:int}), the interaction decays rapidly with distance; thus, the dynamics is effectively dominated by the nearest-neighbor interaction~\cite{olmos2010}.

In the limit $\Delta \gg \Omega, |V_{\ell m}|$, the Rabi oscillation is suppressed because of the large energy gap between the up and down spin states, which suggests that a spin configuration, for example, $| a \rangle = |\cdots \downarrow \downarrow \uparrow
\cdots \rangle$, may be approximately an eigenstate of $H_{\rm R}$. Therefore, given an initial spin configuration, there is no fluctuation in time. However, if $V_{\ell m} = -\Delta$ for the nearest-neighbor pair $\ell$ and $m$, and the $\ell$th spin is up, the $m$th spin can fluctuate with $\Omega$. This is the so-called antiblockade mechanism, in which an excited atom facilitates Rabi oscillation at the nearest-neighbor atom. The effective Hamiltonian can be described in one dimension:
\begin{align}
\label{eq:effective_hamil}
\hat{H}_{\rm{eff}}=\Omega \sum_{\ell} \hat{P}_\ell\hat{\sigma}^x_\ell\,,
\end{align}
where the projection operator is given by $\hat{P}=\hat{n}_{\ell-1}+\hat{n}_{\ell+1}-2\hat{n}_{\ell-1}\hat{n}_{\ell+1}$. 
We are interested in the region near the transition point, where the low-density limit leads to $\hat{P}\approx\hat{n}_{\ell-1}+\hat{n}_{\ell+1}$. Thus, the quantum Hamiltonian in the main text is obtained. Additionally, the decay in the main text is implemented by radiative decay from $\mid\uparrow\rangle$ to $\mid\downarrow\rangle$ using a zero-temperature heat bath.


On the other hand, the classical limit is obtained via the strong dephasing rate, which is denoted as $\Gamma$, as extensively discussed in Ref.~\cite{marcuzzi2015}.
In the strong dephasing limit, it is known that the coherent dynamics can be neglected; thus, the effective Hamiltonian in Eq.~\eqref{eq:effective_hamil} effectively becomes classical and is given by the Lindblad operators in the main text. The strong dephasing limit is derived using superoperator formalism in the next section. The resulting relation is given as $\kappa=4\Omega^2/\Gamma$.

When we combine the quantum and classical settings, the relationship between the experimental realization and QCP is given by $\omega=\Omega_1$ and $\kappa=4\Omega_2^2/\Gamma$, where $\Omega_1$ and $\Omega_2$ are the Rabi frequencies of the independent laser source~\cite{marcuzzi2016,buchhold2017}. 
Thus, for $\Omega_1=\omega_c$ and $\Omega_2<\sqrt{\kappa_*\Gamma}/2$, QDP behavior appears. This region is a quantum critical region where quantum fluctuations play a role in the universal behavior. More details are provided in the next section. 

\section{Realization of classical contact process by Rydberg atom experiment}
\label{appendixB}

The critical behavior of the CCP can be realized using Rydberg atom experiments. Here we show that the transition rate $\kappa$ of the branching and coagulation processes in the classical limit can be obtained in terms of the experimental parameters used in the Rydberg atom system.

\subsection{Lindblad equation for Rydberg gases}

To describe the Rydberg gases in open quantum systems, we employ the Lindblad equation as follows:
\begin{align}
\partial_t\hat{\rho}=
-{\rm i}\left[ \hat{H}_{0}+\hat{H}_{\Omega},\hat{\rho} \right]+ \sum_{i=1}^2 D^{(i)}(\hat{\rho})\,.
\label{eq:Lind_Ryd}
\end{align}
We have split $H_{\rm R}$ into two parts,
\begin{align}
\hat{H}_{0}=\Delta\sum_\ell^N \hat{n}_\ell+\sum_{\ell\neq m}\frac{V_{\ell m}}{2}\hat{n}_\ell \hat{n}_m \quad  
{\rm and}\quad \hat{H}_{\Omega}= \Omega\sum_\ell^N \hat{\sigma}_\ell^x, \nonumber
\end{align}
and the Lindblad dissipator $D^{(i)}$ is given by
\begin{align}
D^{(i)}(\hat{\rho}) =\sum_{\ell} \biggl(\hat{L}^{(i)}_\ell\hat{\rho}\hat{L}^{(i) \dagger}_\ell -\frac{1}{2}\left\{\hat{L}^{(i)\dagger}_\ell \hat{L}^{(i)}_\ell,\hat{\rho} \right\} \biggr).
\label{eq:diss}
\end{align}
Here, the Lindblad operators defined on each site, $\hat{L}^{(1)}_{\ell}$ and $\hat{L}^{(2)}_{\ell}$, represent
decay from $\mid\uparrow\rangle$ to $\mid\downarrow\rangle$  
by a zero-temperature heat bath and the dephasing processes, respectively, where 
the detailed forms are given by
\begin{align}
\hat{L}^{(1)}_{\ell}=\sqrt{\gamma}\hat{\sigma}_\ell^-\quad {\rm and} \quad\hat{L}^{(2)}_{\ell}=\sqrt{\Gamma}\hat{n}_\ell.
\end{align}

By using the representation of the operator with the spin configurations as the basis, for example,
$\rho_{ab} \equiv \langle a |\hat{\rho} |b \rangle$, where $a$ and $b$ denote
one of the $2^N$ configurations of $N$ spins, that is,  
$|a \rangle = \mid \downarrow \uparrow \cdots \downarrow\rangle$,  
each term on the r.h.s. of Eq.~(\ref{eq:Lind_Ryd}) can be evaluated as follows.
Because $\hat{H}_0$ is the diagonal matrix for the basis, the first term reads
\begin{align}
\langle a  | [\hat{H}_{0},\hat{\rho}]  |b \rangle &= 
\Big[ \Delta \sum_{\ell} \left(
\langle a | \hat{n}_{\ell} | a \rangle - \langle b | \hat{n}_{\ell} | b \rangle \right) \nonumber\\
&+ \sum_{\ell \neq m} \frac{V_{\ell m}}{2} \left( 
\langle a | \hat{n}_{\ell} \hat{n}_{m} | a \rangle - 
\langle b | \hat{n}_{\ell} \hat{n}_{m} | b \rangle \right)
\Big]\,\rho_{ab},\nonumber\\
&= \left(
\langle a | \hat{H}_{0} | a \rangle - \langle b | \hat{H}_{0} | b \rangle \right) 
\,\rho_{ab},
\label{eq:H0}
\end{align}
which becomes zero when $a=b$. By contrast, $\hat{H}_{\Omega}$ yields the transition between the
states by flipping a single spin, and the second term in Eq.~(\ref{eq:Lind_Ryd}) can be written as 
\begin{align}
\langle a | [\hat{H}_{\Omega},\hat{\rho}] |b \rangle &=  
\Omega \sum_{\ell} \sum_{a'} \Big[  \langle a | \hat{\sigma}^{x}_{\ell} | a' \rangle \rho_{a'b}
- \langle a' | \hat{\sigma}^{x}_{\ell} | b \rangle \rho_{aa'} \Big],
\label{eq:Ho}
\end{align}
where the diagonal term $a=b$ is given by
\begin{align}
\langle a | [\hat{H}_{\Omega},\hat{\rho}] |a \rangle 
&=  
\Omega \sum_{\ell} \sum_{a'} \Big[  \langle a | \hat{\sigma}^{x}_{\ell} | a' \rangle \rho_{a'a}
- \langle a' | \hat{\sigma}^{x}_{\ell} | a \rangle \rho_{aa'}\Big] \,.
\label{eq:Ho_dia}
\end{align}
Note that the diagonal part of Eq.~(\ref{eq:Ho_dia}) 
contains only the off-diagonal contributions of the density matrix because
$\langle a| \hat{\sigma}^{\pm}_{\ell} |a \rangle =0$, which will be used to derive the rate equation.
Similarly, we obtain the representations of the Lindblad operators; for the decay operator,   
\begin{align}
\langle a | D^{(1)}(\hat{\rho}) |b \rangle &=
\gamma\sum_{\ell}  \sum_{a'b'}  \langle a | \hat{\sigma}^{-}_{\ell} | a' \rangle
\langle b | \hat{\sigma}^{-}_{\ell} | b' \rangle^{*} \rho_{a' b'} \nonumber\\
& - \frac{\gamma}{2} \sum_{\ell} 
 \left( \langle a| \hat{n}_{\ell} | a \rangle  +  
\langle b| \hat{n}_{\ell} | b \rangle  \right)  \rho_{ab} \,,
\label{eq:D1}
\end{align} 
where the diagonal term $a=b$ is given by
\begin{align}
\langle a | D^{(1)}(\hat{\rho}) |a \rangle &=
\gamma\sum_{\ell}\sum_{a'}   \Big[   |\langle a | \hat{\sigma}^{-}_{\ell} | a' \rangle|^2 \rho_{a' a'} - 
 |\langle a'| \hat{\sigma}^-_{\ell} | a \rangle|^2    \rho_{aa} \Big]\,.
\end{align} 
For the dephasing operator,
\begin{align}
\langle a | D^{(2)}(\hat{\rho}) |b \rangle &= 
\Gamma\sum_{\ell} \left(  \langle a| \hat{n}_{\ell} | a \rangle \langle b| \hat{n}_{\ell} | b \rangle
- \frac{\langle a| \hat{n}_{\ell} | a \rangle}{2} - \frac{\langle b| \hat{n}_{\ell} | b \rangle}{2}                 
\right) \rho_{ab} \nonumber\\
&= -\frac{\Gamma}{2} \sum_{\ell}|\langle a| \hat{n}_{\ell} | a \rangle
-\langle b| \hat{n}_{\ell} | b \rangle | \,\rho_{ab}\,,
\label{eq:D2}
\end{align}
which becomes zero when $a = b$.
If the Lindblad equation consists of only this dephasing dissipator,
the density operator $\rho_{ab}$ in the long time limit becomes the diagonal matrix; therefore, $D^{(2)}$ is called the dephasing operator. It is known that in the limit $\Gamma\gg\Omega, \gamma$, 
the coherent dynamics can be neglected, and thus 
the Lindblad equation [Eq.~(\ref{eq:Lind_Ryd})] effectively reduces to 
the classical rate equation~\cite{marcuzzi2015}.
We will briefly show the procedure below. 

\subsection{Derivation of transition rate in classical limit}
For convenience, we introduce the superoperator~\cite{uzdin2015} by mapping the 
density operator to the density vector, $\rho_{ab} \to \rho_{\alpha}$, where $a$ and $b$ are those of 
the spin configurations as defined above, 
and $\alpha$ is the corresponding vector index. Then we rewrite the 
Lindblad equation [Eq.~(\ref{eq:Lind_Ryd})] in terms of the vectorized density operator $\vec{\rho}$ as
\begin{align}
\partial_t \vec{\rho} = -{\rm i} \hat{\mathcal{H}} \vec{\rho},
\label{eq:Lind_vec}
\end{align} 
where the superoperator $\hat{\mathcal{H}}$ defined in
a $4^N \times 4^N$ complex space is given by
\begin{align}
\mathcal{H}_{\alpha \beta} \equiv {\rm i} \frac{ \partial 
\left( \partial_t \rho_{\alpha} \right)} {\partial \rho_{\beta}} 
\end{align}
because Eq.~(\ref{eq:Lind_Ryd}) is the linear equation of $\rho_{\alpha}$.
We decompose the density vector to two parts as 
$\vec{\rho} = \vec{\mu} \oplus \vec{\nu}$, where $\vec{\mu}$ (which belongs to the $2^N$-dimensional space $\mathcal{M}$) and $\vec{\nu}$ (which belongs to the $4^N-2^N$-dimensional space $\mathcal{N}$)
are defined by arranging $\rho_{ab}$ as follows:
\begin{align}
\vec{\mu} &= ( \underbrace{ \rho_{0}, \rho_{1}, \cdots ,\rho_{\alpha}, \cdots, \rho_{2^N-1} }_{
\rho_{\alpha} =\rho_{aa}}
)^{T} \in \mathcal{M} \,,\nonumber\\ 
\vec{\nu} &= ( \underbrace{ \rho_{2^N}, \rho_{2^N +1}, \cdots ,\rho_{\alpha}, \cdots \rho_{4^N-1} }_{ 
\rho_{\alpha}= \rho_{a\neq b}} )^{T} \in \mathcal{N}\,.
\end{align} 
Here, the first $2^N$ components of the vectors correspond to the diagonal components of the density matrix,
and the others are the off-diagonal terms of the density matrix. 
From these decomposed vectors and Eqs.
~(\ref{eq:H0})--(\ref{eq:D2}), $\mathcal{\hat{H}}$ can be decomposed using the block matrix
\begin{align}
\label{eq:matrix}
\mathcal{\hat{H}}=\left(\begin{array}{c|c} 
\mathcal{\hat{H}}^{(1)}_1 & \mathcal{\hat{H}}_1^{\Omega} \\ 
\hline
\mathcal{\hat{H}}_2^{\Omega} & \mathcal{\hat{H}}^{0}+\mathcal{\hat{H}}_3^{\Omega}+\mathcal{\hat{H}}^{(1)}_2+\mathcal{\hat{H}}^{(2)} \\ 
\end{array}\right),
\end{align}
where the upper left (lower right) part of the matrix is mapped to $\mathcal{M}\to\mathcal{M}$ ($\mathcal{N}\to\mathcal{N}$), and the upper right (lower left) is mapped to $\mathcal{M}\to\mathcal{N}$ ($\mathcal{N}\to\mathcal{M}$). In addition, $\hat{\mathcal{H}}^{(1)}=\hat{\mathcal{H}}^{(1)}_1+\hat{\mathcal{H}}^{(1)}_2$ is separable into two parts: $\hat{\mathcal{H}}^{(1)}_1: \mathcal{M} \to \mathcal{M}$ and $\hat{\mathcal{H}}^{(1)}_2: \mathcal{N} \to \mathcal{N}$. Likewise, $\hat{\mathcal{H}}^{\Omega}$ is separable into three parts: $\hat{\mathcal{H}}^{\Omega}_1: \mathcal{M} \to \mathcal{N}$, $\hat{\mathcal{H}}^{\Omega}_2: \mathcal{N} \to \mathcal{M}$, and $\hat{\mathcal{H}}^{\Omega}_3: \mathcal{N} \to \mathcal{N}$. Further, $\hat{\mathcal{H}}^{\Omega}_1$ and $\hat{\mathcal{H}}^{\Omega}_2$ are switching operators that switch between the spaces of $\mathcal{M}$ and $\mathcal{N}$ by flipping a single spin.
The components of $\alpha=ab$ for the block matrix in Eq.~\eqref{eq:matrix} are defined as follows: 
\begin{align}
\label{eq:superH1}
\mathcal{H}^{0}_{\alpha \alpha} &= \frac{\partial \left( [\hat{H}_{\Omega}, \hat{\rho}] \right)_{\alpha}}
{\partial \rho_{\alpha}}= \langle a | \hat{H}_{0} | a \rangle - \langle b | \hat{H}_{0} | b \rangle    ,\\
\label{eq:superH4}
\mathcal{H}^{(2)}_{\alpha \alpha} &= {\rm i}\frac{\partial \left( D^{(2)}(\hat{\rho}) \right)_{\alpha}}
{\partial \rho_{\alpha}}= -\frac{{\rm i}\Gamma}{2} \sum_{\ell}|\langle a| \hat{n}_{\ell} | a \rangle
-\langle b| \hat{n}_{\ell} | b \rangle | \,,\\
\label{eq:superH2}
\mathcal{H}^{\Omega}_{1~\alpha \beta} &= \frac{\partial \left( [\hat{H}_{\Omega}, \hat{\rho}] \right)_{\alpha}}
{\partial \rho_{\beta}}\nonumber\\
&=\Omega \sum_{\ell} \sum_{a'} \Big[  \langle a | \hat{\sigma}^{x}_{\ell} | a' \rangle \delta_{\beta,a'a}
- \langle a' | \hat{\sigma}^{x}_{\ell} | a \rangle \delta_{\beta,aa'}\Big] \,, \\
\label{eq:superH2_2}
\mathcal{H}^{\Omega}_{2~\alpha \beta} &= \frac{\partial \left( [\hat{H}_{\Omega}, \hat{\rho}] \right)_{\alpha}}
{\partial \rho_{\beta}}\nonumber\\
&=\Omega \sum_{\ell} \Big[  \langle a | \hat{\sigma}^{x}_{\ell} | b \rangle \delta_{\beta,bb} 
- \langle a | \hat{\sigma}^{x}_{\ell} | b \rangle \delta_{\beta,aa}\Big] \,, \\
\label{eq:superH3}
\mathcal{H}^{(1)}_{1~\alpha \beta} &= {\rm i}\frac{\partial \left( D^{(1)}(\hat{\rho}) \right)_{\alpha}}
{\partial \rho_{\beta}}\nonumber\\
&=
{\rm i}\gamma\sum_{\ell}   \sum_{a'} \Big[ |\langle a | \hat{\sigma}^{-}_{\ell} | a' \rangle|^2 \delta_{\beta,a' a'} - 
 |\langle a'| \hat{\sigma}^-_{\ell} | a \rangle|^2    \delta_{\beta,aa} \Big]\,, 
\end{align}
where each $\left( \cdot \right)_{\alpha}$ denotes the matrix element defined in Eqs.~\eqref{eq:H0}–-\eqref{eq:D2} corresponding to the vector index $\alpha$.

Consequently, one can rewrite Eq.~\eqref{eq:Lind_vec} in terms of $\vec{\mu}$ and $\vec{\nu}$:    
\begin{align}
\label{eq:Lind_vec2}
\partial_{t}\vec{\mu} &= -{\rm i} \hat{\mathcal{H}}^{(1)}_1 \vec{\mu} -{\rm i} \hat{\mathcal{H}}^{\Omega}_{1} \vec{\nu} \,,
\\
\label{eq:Lind_vec3}
\partial_{t}\vec{\nu} &= -{\rm i} \hat{\mathcal{H}}^{\Omega}_{2} \vec{\mu}-{\rm i} \left( \hat{\mathcal{H}}^{0} + \hat{\mathcal{H}}^{\Omega}_{3} 
+\hat{\mathcal{H}}^{(1)}_2 +\hat{\mathcal{H}}^{(2)} \right) \vec{\nu}
\,.
\end{align}

In the strong dephasing limit, where $\hat{\mathcal{H}}^{(2)}$ 
may dominate the off-diagonal dynamics in Eq.~(\ref{eq:Lind_vec3}),
the solution of Eq.~\eqref{eq:Lind_vec3} becomes an approximately exponentially decaying function 
in time with a time scale of $1/\Gamma$ ($\Delta$ is also a large parameter, but 
$\hat{\mathcal{H}}^{0}$ only induces oscillation).
Therefore, the full dynamics can be effectively reduced to the diagonal dynamics of $\vec{\mu}$
at a slower time scale than $1/\Gamma$.     
Inserting the solution of Eq.~(\ref{eq:Lind_vec3}), which is given by
\begin{align}
\vec{\nu}(t) &= e^{-{\rm i}\left( \hat{\mathcal{H}}^{0} + \hat{\mathcal{H}}^{\Omega}_{3}
+\hat{\mathcal{H}}^{(1)}_2 +\hat{\mathcal{H}}^{(2)} \right)t}\, \vec{\nu}(0)\nonumber\\
&-{\rm i}\int_0^tdt'\,e^{-{\rm i}\left( \hat{\mathcal{H}}^{0} + \hat{\mathcal{H}}^{\Omega}_{3}+
\hat{\mathcal{H}}^{(1)}_2 +\hat{\mathcal{H}}^{(2)} \right)(t-t')}  \hat{\mathcal{H}}^{\Omega}_{2} \vec{\mu}(t'),
\end{align}
into Eq.~(\ref{eq:Lind_vec2}) with $\vec{\nu}(0)=0$, we obtain
\begin{align}
\partial_{t}\vec{\mu} &= -{\rm i} \hat{\mathcal{H}}^{(1)}_1 \vec{\mu}(t)  \nonumber\\
&-\int_{0}^t dt'\, \hat{\mathcal{H}}^{\Omega}_{1}e^{-{\rm i}\left( \hat{\mathcal{H}}^{0} + \hat{\mathcal{H}}^{\Omega}_{3}+
\hat{\mathcal{H}}^{(1)}_2 +\hat{\mathcal{H}}^{(2)} \right)(t-t')}
\hat{\mathcal{H}}^{\Omega}_{2} \vec{\mu}(t').
\label{eq:rate_1}
\end{align}
Because $\Gamma, \Delta \gg \gamma, \Omega$, we can expand the exponential function about the small parameter using the Zassenhaus formula~\cite{sridhar2003}; it then becomes
\begin{align}
\partial_{t}\vec{\mu} &= -{\rm i} \hat{\mathcal{H}}^{(1)}_1 \vec{\mu}(t)  
-\int_0^t dt'\, \hat{\mathcal{H}}^{\Omega}_{1}e^{-{\rm i}\left( \hat{\mathcal{H}}^{0} 
+\hat{\mathcal{H}}^{(2)} \right)(t-t')}
\hat{\mathcal{H}}^{\Omega}_{2} \vec{\mu}(t')\nonumber\\
&+\mathcal{O}(\Omega^2\gamma,\Omega^3).\label{eq:rate_2}
\end{align}
We remark again that the time scale for the dynamics of $\vec{\mu}$ is much larger than $1/\Gamma$, which results in the replacement $\vec{\mu}(t') \to \vec{\mu}(t)$ in the integrand of Eq.~(\ref{eq:rate_2}), as in the Markov approximation~\cite{breuer2002}.
In addition, we let the lower limit of the integral go to negative infinity because of the rapid decay of the exponential function. Then, rearranging the time integral in Eq.~(\ref{eq:rate_2}) as $(t-t') \to t'$, we obtain
\begin{align}
\partial_{t}\vec{\mu} \approx -{\rm i} \hat{\mathcal{H}}^{(1)}_1 \vec{\mu}(t)  
-\int_0^\infty dt'\, \hat{\mathcal{H}}^{\Omega}_{1}e^{-{\rm i}\left( \hat{\mathcal{H}}^{0} 
+\hat{\mathcal{H}}^{(2)} \right)t'}
\hat{\mathcal{H}}^{\Omega}_{2} \vec{\mu}(t)\,.
\label{eq:rate}
\end{align}
Now, one can see that the slow dynamics in Eq.~(\ref{eq:rate}) is given only in terms of $\mu_{\alpha}$,
and thus it can be rewritten using the diagonal elements $\rho_{aa}$ of the density operator. 
The first term on the r.h.s. of Eq.~(\ref{eq:rate}) is given by the diagonal element of Eq.~\eqref{eq:superH3}, which is composed only of diagonal elements 
$\mu_{\alpha}$ or $\rho_{aa}$ as follows:
\begin{align*}
\bigg(-{\rm i} \hat{\mathcal{H}}^{(1)}_1 \vec{\mu}(t)\bigg)_\alpha
=\gamma\sum_{\ell}\sum_{b}  \Big[  |\langle a | \hat{\sigma}^{-}_{\ell} | b \rangle|^2 \rho_{bb} -. 
|\langle b | \hat{\sigma}^{-}_{\ell} | a \rangle|^2\rho_{aa} \Big]\,.
\end{align*}
Next, because $\hat{\mathcal{H}}^{0}$ and $\hat{\mathcal{H}}^{(2)}$ are diagonal matrices, the $\alpha$th component of the second term is given by
\begin{align}
&\bigg( -\int_0^\infty dt'\, \hat{\mathcal{H}}^{\Omega}_{1}e^{-{\rm i}\left( \hat{\mathcal{H}}^{0}
+\hat{\mathcal{H}}^{(2)} \right)t'}
\hat{\mathcal{H}}^{\Omega}_{2} \vec{\mu} \bigg)_{\alpha}\nonumber\\=
&-\sum_{\beta ,\alpha'} \mathcal{H}^{\Omega}_{1~\alpha \beta}
\mathcal{H}^{\Omega}_{2~ \beta \alpha'} \mu_{\alpha'} 
\times\int_0^{\infty} dt' e^{-{\rm i}\left( \mathcal{H}^{0}_{\beta \beta}
+\mathcal{H}^{(2)}_{\beta \beta} \right)t'}.
\label{eq:rate2}
\end{align}
Using Eqs.~\eqref{eq:superH2} and~\eqref{eq:superH2_2}, one can rewrite 
Eq.~\eqref{eq:rate2} by changing $\alpha \to aa$: 
\begin{align}
&\Omega^2
\sum_{b} \left( \rho_{bb} -\rho_{aa} \right) 
\sum_{\ell} \left| \langle a | \hat{\sigma}^{x}_\ell | b \rangle \right|^2 \nonumber\\
& \quad \quad \times \int_0^{\infty} dt' \left( e^{-{\rm i}\left( \mathcal{H}^{0}_{ab,ab }, 
+\mathcal{H}^{(2)}_{ab,ab} \right)t'} 
+e^{-{\rm i}\left( \mathcal{H}^{0}_{ba,ba }
+\mathcal{H}^{(2)}_{ba,ba} \right)t'} \right) \nonumber\\
&\equiv \sum_{b} \left( \Lambda_{a,b} \rho_{bb}-\Lambda_{b,a} \rho_{aa} \right) \,,
\label{eq:rate3}
\end{align}
where we have used $\sum_{\ell,\ell'} \left( \langle a | \hat{\sigma}^{x}_\ell | b \rangle \right)
\left( \langle b | \hat{\sigma}^{x}_{\ell'}  | a \rangle \right) =\sum_{\ell} \left| \langle a | \hat{\sigma}^{x}_\ell | b \rangle \right|^2$, which becomes zero when $\ell$ and $\ell'$ are different. Note that the transition rate $\Lambda_{a,b}=\Lambda_{b,a}$ induced by the interaction
is nonzero only when the spin configuration $b$ is given by $|b \rangle=|a_1 \rangle$,
where $\{|a_1 \rangle\}$ is generated from $|a\rangle$ 
by flipping a single spin. For example, 
if $|a \rangle = \left| \uparrow \downarrow \downarrow \right\rangle$ is given for $N=3$, 
a set $\{ |a_1 \rangle \}$ can be defined as $\{  \left| \downarrow \downarrow \downarrow \right\rangle ,
\left| \uparrow \uparrow \downarrow \right\rangle, \left| \uparrow \downarrow \uparrow \right\rangle \}$.
Now, using $\mathcal{H}^{(2)}_{aa_1,aa_1} = -{\rm i}\Gamma/2$ for any $a_1$
and $\mathcal{H}^{0}_{aa_1,aa_1} = \langle a| \hat{H}_{0} | a \rangle
- \langle a_1| \hat{H}_{0} | a_1 \rangle $,
we obtain the nonzero transition rate $\Lambda_{a,a_1}$ in Eq.~(\ref{eq:rate3}), as follows: 
\begin{eqnarray}
\Lambda_{a,a_1}&=&\Omega^2 
\int_{0}^{\infty}dt' 2e^{-\frac{\Gamma}{2} t'}\cos \left(  \left\{ \langle a| \hat{H}_{0} | a \rangle 
- \langle a_1| \hat{H}_{0} | a_1 \rangle
\right \} t' \right)  \nonumber\\
&=&   \frac{\Gamma \Omega^2}{ \left(\Gamma/2 \right)^2 + 
\left( \langle a| \hat{H}_{0} | a \rangle - \langle a_1| \hat{H}_{0} | a_1 \rangle \right)^2 }
\,.
\label{eq:rate4}
\end{eqnarray}
From Eqs.~\eqref{eq:D1} and~\eqref{eq:rate3}, therefore, one can see that the rate equation 
Eq.~\eqref{eq:rate} up to the second order is written as
\begin{align}
\partial_t \rho_{aa} = \sum_{b} \left( W_{a,b} \rho_{bb} -  W_{b,a} \rho_{aa} \right) \,,
\label{eq:rate5}
\end{align}
where the transition rate $W_{a,b}$ reads
\begin{align}
\label{eq:rate6}
W_{a,b}=\gamma \sum_{\ell} |\langle a| \hat{\sigma}^{-}_\ell | b \rangle|^2
+ \Lambda_{a,b}.
\end{align}

\subsection{DERIVATION OF $\kappa=4\Omega^2/\Gamma$}
\label{sec:rate}
First, we consider only the nearest-neighbor interaction, $V_{\ell m}= V_0$, for the nearest-neighbor pair $(\ell ,m)$; otherwise, $V_{\ell m}=0$. Defining $a_{\ell}$ as 
$|a_{1\ell} \rangle =  \hat{\sigma}_\ell^{x} | a \rangle$, we obtain   
the transition rate $\Lambda_{a,a_\ell}$ in Eq.~\eqref{eq:rate4} as
\begin{align}
\Lambda_{a, a_{1\ell}} 
=\frac{4\Omega^2}{\Gamma}\frac{1}{ 1 +\frac{4}{\Gamma^2} 
\left(\Delta + \langle a| \hat{P}_{\ell} |a\rangle V_0 \right)^2 } \,,
\label{eq:nnrate}
\end{align}
where $\hat{P}_{\ell}$ denotes the number of nearest neighbors for which the site $\ell$ is in the up state,
that is, 
$\hat{P}_{\ell} \equiv \sum_{m\in {\rm n.n.}(\ell)} \hat{n}_{m}$.
Setting $V_0= -\Delta$ and $\Delta \gg \Gamma$, one can see that
Eq.~\eqref{eq:nnrate} is approximately zero, except for 
$\langle a| \hat{P}_{l} |a \rangle=1$. In the low-density limit,  
where the number of up spins per site is vanishingly small 
($n={\rm tr} \left[ \hat{\rho} \sum_{\ell} \hat{n}_{\ell} \right]/N \ll 1$),
the configurations having a small number of up spins make a major 
contribution to $\hat{\rho}$.  
Then one can assume that $\langle a| \hat{P}_{\ell} |a \rangle$ for
the major configuration is generally 0 or 1;  
thus, we obtain the following approximation:  
\begin{align}
\Lambda_{a, a_\ell} \approx 
\frac{4 \Omega^2} {\Gamma} \langle a| \hat{P}_{\ell} |a \rangle
=\sum_{m\in{\rm n.n}(\ell)} \frac{4 \Omega^2} {\Gamma} \langle a| \hat{n}_m |a \rangle \,.
\end{align}
Using $|a \rangle = \hat{\sigma}^x_\ell | a_{1\ell} \rangle$ and considering general configurations $b$,
the transition rate $\Lambda_{a,b}$ becomes
\begin{align}
\Lambda_{a,b}
=\frac{4\Omega^2}{\Gamma}\sum_{\ell}\sum_{ m\in {\rm n.n.}(\ell)} 
\left( \left| \langle a| \hat{n}_m \hat{\sigma}^+_\ell  |b \rangle \right|^2
+ \left| \langle a| \hat{n}_m \hat{\sigma}^-_\ell |b \rangle \right|^2 \right) \,,
\label{eq:nnrate2}
\end{align}
which is equivalent to the branching and coagulation processes in 
the ordinary CP model. 
Here we used $\langle a| \hat{n}_m \hat{\sigma}^+_\ell |b \rangle + 
\langle a| \hat{n}_m \hat{\sigma}^-_\ell |b \rangle
=|\langle a| \hat{n}_m \hat{\sigma}^+_\ell |b \rangle|^2 
+ |\langle a|  \hat{n}_m \hat{\sigma}^-_\ell |b \rangle|^2$
in Eq.~\eqref{eq:nnrate2}.
We expect that this approximation in the low-density limit
may be valid near the absorbing transition point, where the order parameter $n$ is small. 

Finally, we briefly present the diagonal component of the Lindblad equation with the Lindblad operator $\hat{L}_\ell$ corresponding to Eq.~\eqref{eq:rate5}:
\begin{align}
\partial {\rho}_{aa} &=\langle a|\sum_{\ell} \Big(\hat{L}_\ell\hat{\rho}\hat{L}^{\dagger}_\ell -\frac{1}{2}\Big\{\hat{L}^{\dagger}_\ell \hat{L}_\ell,\hat{\rho} \Big\} \Big) |a\rangle \nonumber\\
&=\sum_{b} \Big(\sum_{\ell} |\langle a|\hat{L}_\ell|b\rangle|^2 \rho_{bb} 
-\sum_{\ell} |\langle b|\hat{L}_\ell|a\rangle|^2 \rho_{aa}\Big)\,,
\end{align}
where the transition rate is given by $W_{a,b}=\sum_{\ell} |\langle a|\hat{L}_\ell|b\rangle|^2$. Thus, from Eqs.~\eqref{eq:rate6} and~\eqref{eq:nnrate2}, the three Lindblad operators are obtained with $\kappa=4\Omega^2/\Gamma$.
\section{Classical contact process using the quantum jump Monte Carlo method}
\label{appendixC}
\begin{figure}[h!]
\includegraphics[width=1.00\columnwidth]{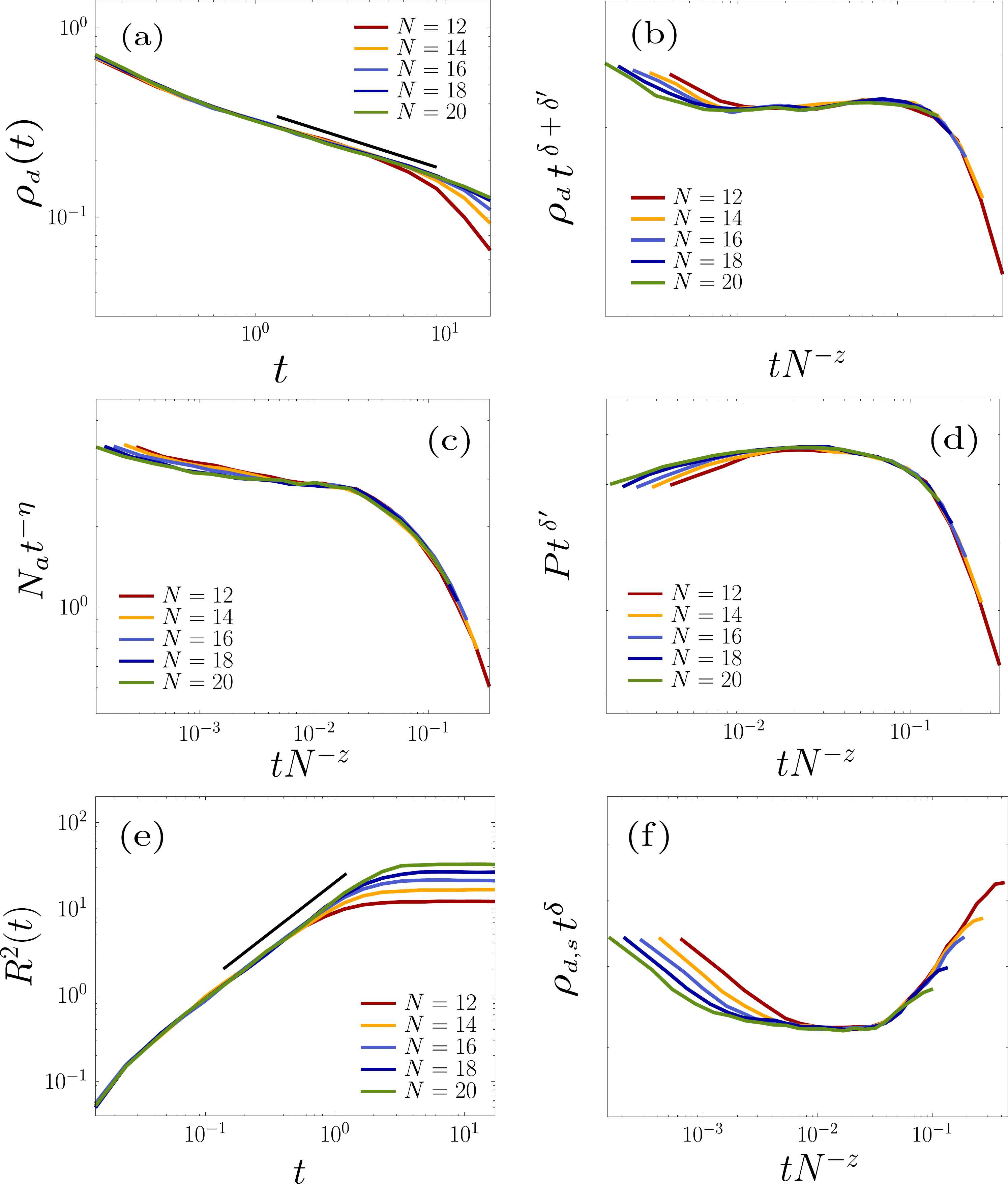}
\caption{Estimates of the critical exponents of the 1D-CCP starting from a single active initial state. 
(a) Plot of $\rho_d(t)$ versus $t$, which behaves as $\rho_d(t)\sim t^{-\delta-\delta'}$. The solid line is a guideline with slope $-0.32$. Inset: scaling plot of $\rho_d(t)t^{\delta+\delta'}$ versus $tN^{-z}$ for $\delta+\delta'=0.32$ and $z=1.58$.
(b) Scaling plot of $N_a(t) t^{-\eta}$ versus $tN^{-z}$ for $\eta=0.30$ and $z=1.58$.
(c) Scaling plot of $P(t)t^{\delta'}$ versus $tN^{-z}$ for $\delta' =0.16$ and $z=1.58$.
(d) Plot of $R^2(t)$ as a function of $t$. The solid line is a guideline with slope $2/z$ for $z=1.58$. 
(e) Scaling plot of $\rho_{d,s}(t)t^{\delta}$ versus $tN^{-z}$ for $\delta=0.16$ and $z=1.58$. The parameter $t$ is given in units of $1/\gamma$.
\label{fig:fig9}}
\end{figure}

\begin{figure}[ht!]
\includegraphics[width=1.00\columnwidth]{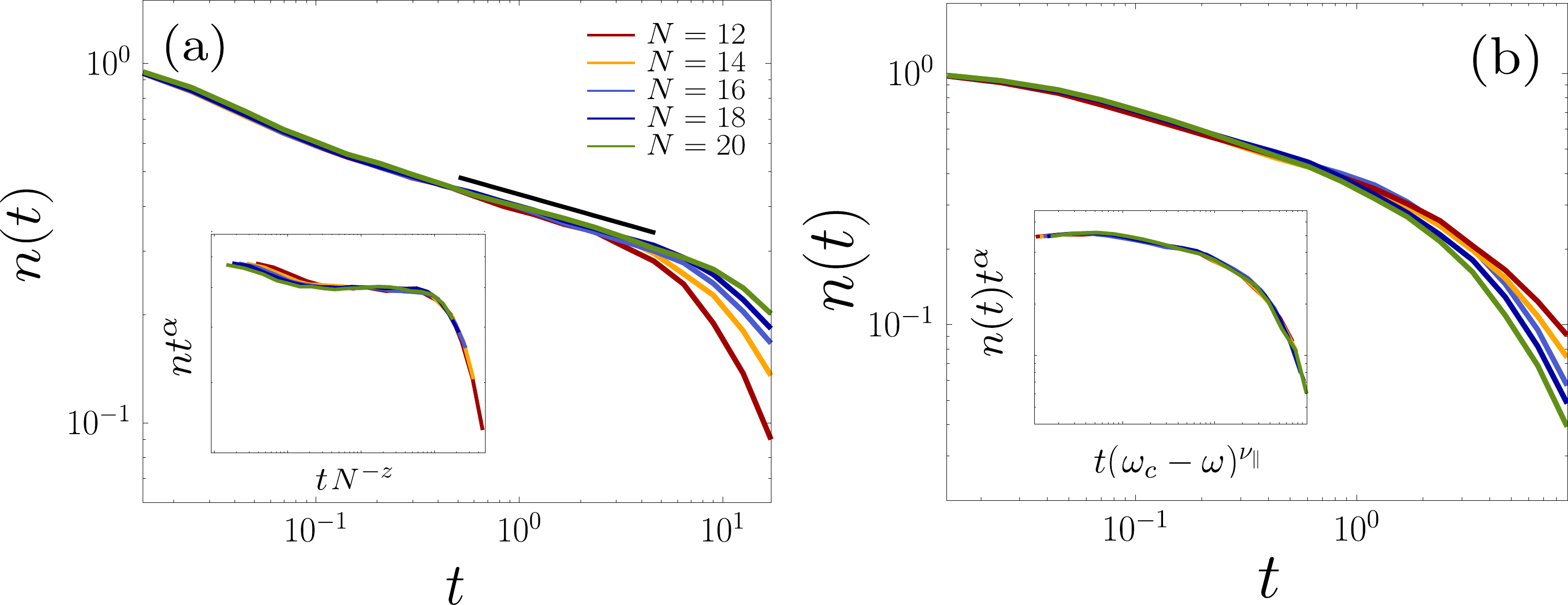}
\caption{Estimates of the critical exponents of the 1D-CCP starting from the fully active initial state. (a) Plot of $n(t)$ as a function of $t$, which shows $n(t)\sim t^{-\alpha}$. The solid line is a guideline with slope $-0.16$. Inset: scaling plot of $n(t)t^{\alpha}$ versus $tN^{-z}$ for $\alpha=0.16$ and $z=1.58$.
(b) Plot of $n(t)$ as a function of $t$ for different values of $\omega<\omega_c$. 
Inset: Data points collapse well onto a single curve for $\alpha=0.16$ and $\nu_{\|}=1.73$. The parameter $t$ is given in units of $1/\gamma$.
\label{fig:fig10}}
\end{figure}

\begin{figure*}[!t]
\includegraphics[width=2.00\columnwidth]{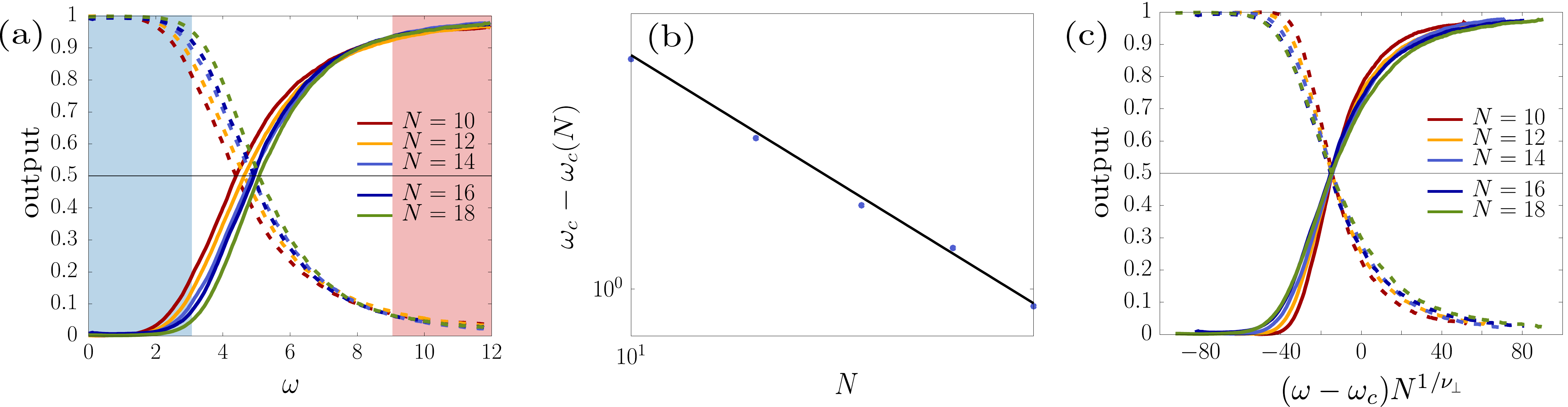}
\caption{Plots obtained using NN approach. (a) Plot of the output averaged over a test set as a function of $\omega$ for different system sizes. The value of the first (second) output neuron is represented as a solid (dashed) line. From this plot, we estimate the crossing point of the two outputs and regard it as the transition point $\omega_c(N)$ for a given system size $N$. The shaded regions $\omega\in [0,3]$ and $\omega\in [9,12]$ indicate the training sets used in the convolutional NN analysis. (b) Plot of $\omega_c-\omega_c(N)$ versus $N$, where $\omega_c$ is chosen to yield power-law behavior, which is typical near the transition point $\omega_c$. The slope represents the value of the critical exponent $-1/{\nu_{\bot}}$. (c) Scaling plot of the output versus $(\omega-\omega_c)N^{1/\nu_{\bot}}$. For the obtained numerical values of $\nu_{\bot}$ and $\omega_c$, the data collapse well for system sizes $N = 10, 12, 14, 16$, and $18$. From (b) and (c), we obtain $\omega_c\approx 6.04$ and $\nu_{\bot}=1.06\pm0.04$. The control parameter is given in units of $\gamma$.} 
\label{fig:fig11}
\end{figure*}

\begin{figure}[h!]
	\centering
    \includegraphics[width=0.8\linewidth]{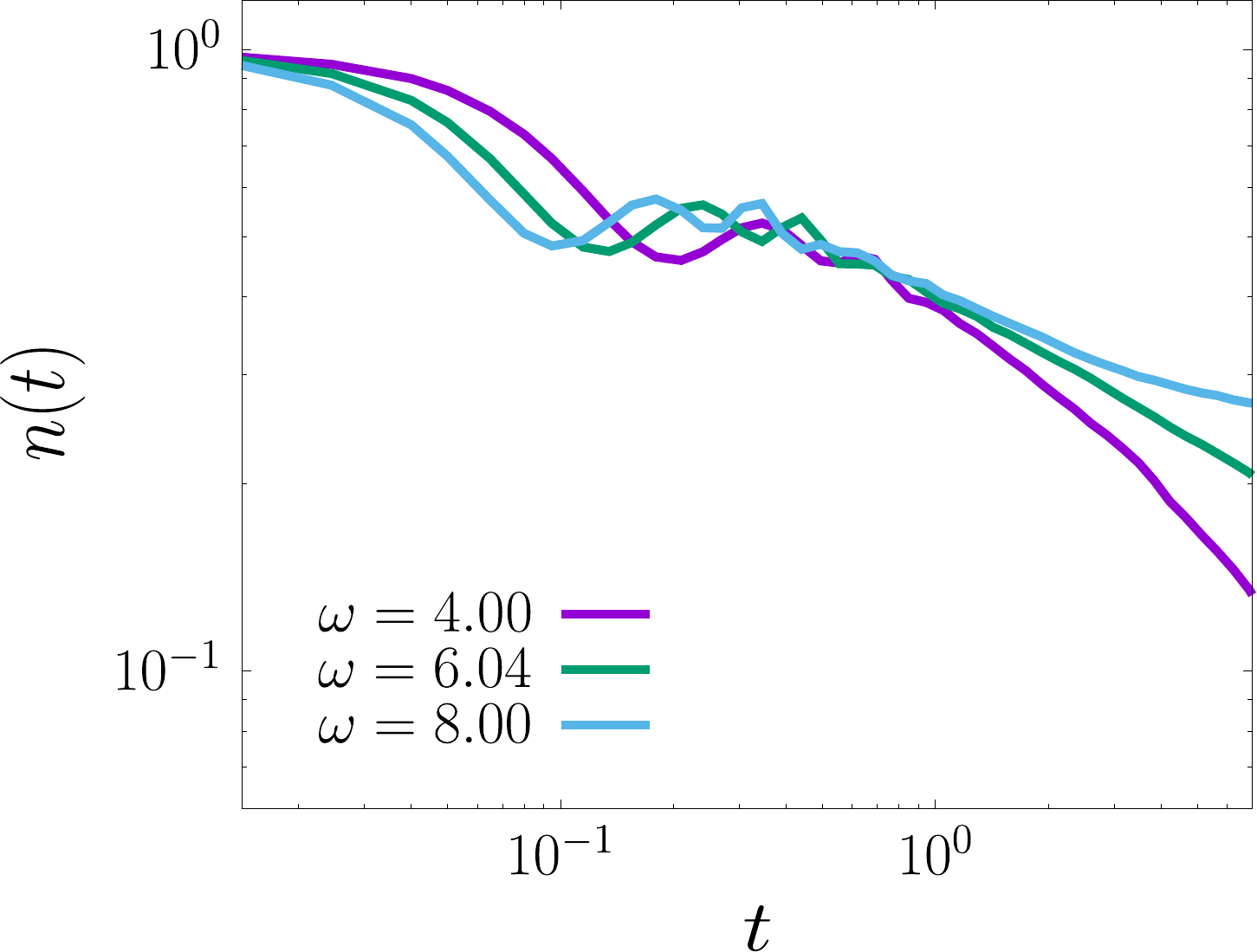}
	\caption{Plot of $n(t)$ as a function of $t$ for different $\omega$.
	For $\omega=4.00$, an exponentially decaying curve is observed. By contrast, for $\omega=8.00$, a stationary state converges to a finite density.
At the critical point $\omega=6.04$, it exhibits power-law behavior. The system size is $N=20$.}
\label{fig:fig12}
\end{figure}

\begin{figure}[!h]
\includegraphics[width=1.00\columnwidth]{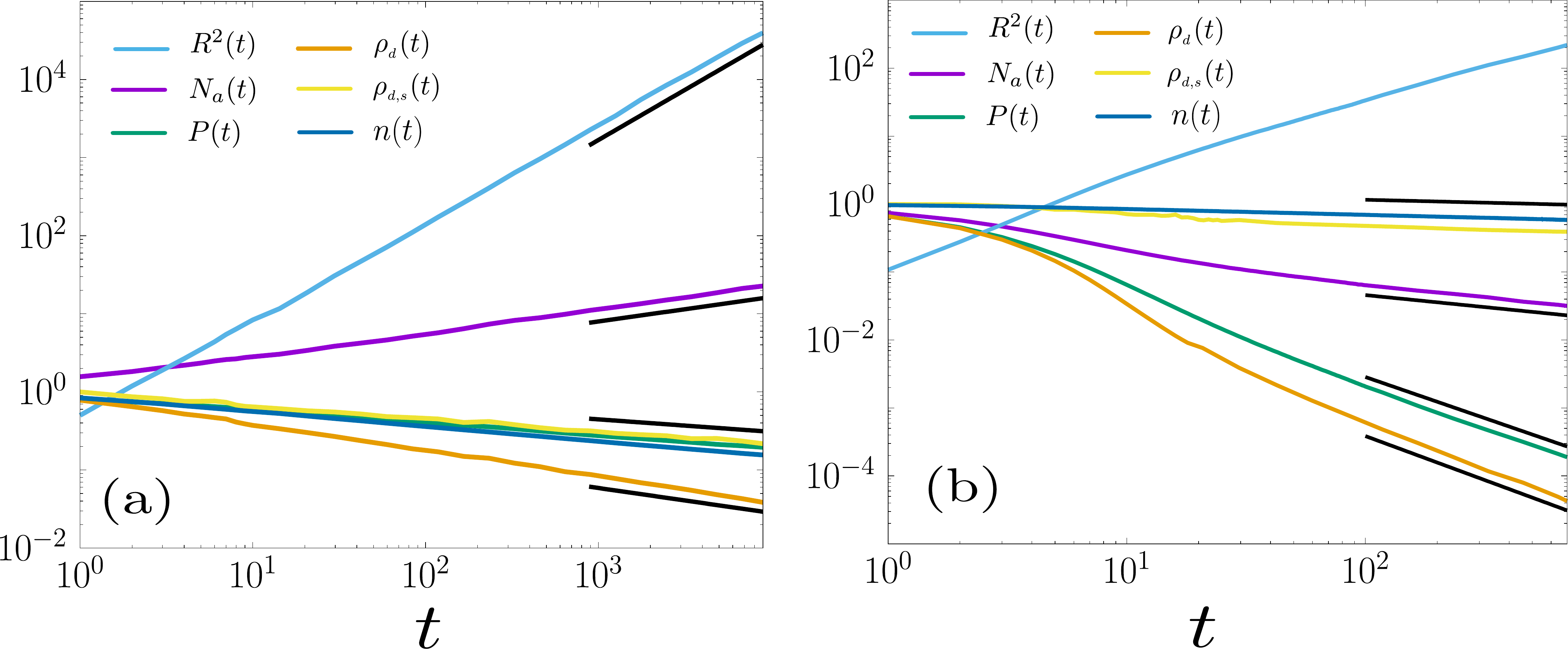}
\caption{Behaviors of physical quantities at the transition point as a function of $t$. 
(a) For the CCP. The solid lines are guidelines with slope $2/z$, $\eta$, $-\delta=-\delta'=-\alpha$, and $\eta-1/z$, from top to bottom. The critical exponents are $z=1.58$, $\delta=0.16$, $\eta=0.31$, $\delta'= 0.16$, and $\alpha=0.16$. Note that $\delta'=\alpha$, and rapidity-reversal symmetry holds.
(b) For the 2D classical tricritical CP starting from a single active site. The solid lines are guidelines with slope $2/z$, $-\delta=-\alpha$, $\eta$, $-\delta'$, and $\eta-1/z$, from top to bottom. The critical exponent values are $z=2.11$, $\delta=0.09$, $\eta=-0.35$, and $\delta'= 1.21$. Note that rapidity-reversal symmetry is broken.} 
\label{fig:fig13}
\end{figure}

In this section, we consider $\omega\to 0$. When $\kappa\ll \gamma$, inactive particles become more abundant with time, and ultimately the system is fully occupied by inactive particles. Thus, the system is no longer dynamic and falls into an absorbing state. When $\kappa \gg \gamma$, the system remains in an active state with a finite density of active particles. Thus, the CCP exhibits a phase transition from an active to an absorbing state as $\kappa$ is decreased.

The critical exponents of the 1D-QCP were obtained using FSS of the data obtained using the QJMC method as described in the main text. 
To check the validity of the FSS for a small system, we consider the 1D-CCP, where $\kappa$ is finite, and $\omega=0$ [see Eqs. (2--5) in the main text]. 
At the critical point, we perform FSS of the 1D-CCP using the QJMC method. The observables are those defined in the main text.

First, we obtain the exponents $\delta+\delta'$, $\eta$, $\delta'$, $z$, $\delta$, and $\alpha$ directly by measuring the slopes of the double-logarithmic plots shown in Figs.~\ref{fig:fig9} and~\ref{fig:fig10}. Then we collapse the data by using the obtained exponents to compute the dynamic exponent $z$.
Specifically, we plot $\rho_d t^{\delta+\delta'}$ versus $tN^{-z}$ in Fig.~\ref{fig:fig9}(a), $N_at^{-\eta}$ versus $tN^{-z}$ in Fig.~\ref{fig:fig9}(b), and $P(t)t^{-\delta'}$ versus $tN^{-z}$ in Fig.~\ref{fig:fig9}(c) for different system sizes $N$.
We measure the exponent $z$ directly using the plot of $R^2(t)$ versus $t$ in Fig.~\ref{fig:fig9}(d). In the CCP, we can classify the surviving runs, and thus we measure the exponent $-\delta$ directly using the plot of $\rho_{\text{d,s}}(t)$ versus $t$ in Fig.~\ref{fig:fig9}(e). 
Next, we plot $n(t)t^{-\alpha}$ versus $tN^{-z}$ in Fig.~\ref{fig:fig10}(a) for different system sizes $N$. 
The exponent $\nu_{\|}$ is obtained from the rescaling plot of $n(t)t^{\alpha}$ versus $t(\omega_c-\omega)^{\nu_{\|}}$ for different $\omega$ in Fig.~\ref{fig:fig10}(b).

The critical exponents are thus obtained as $\delta+\delta'=0.32\pm 0.01$, $\eta=0.31 \pm 0.02$, $\delta'= 0.16\pm 0.01$, $\delta=0.16\pm 0.02$, $z=1.58 \pm 0.03$, and $\alpha=0.16\pm 0.01$. Note that $\delta=\alpha$. In addition, $\alpha=\delta'$, indicating that rapidity-reversal symmetry holds. All critical exponents are in good agreement with the DP values within the error bars.
Thus, we verified that the critical exponents in the CCP can be successfully obtained using the QJMC method with the same system size as in the main text.

\section{Critical behavior obtained by neural network approach with different training regions}
\label{appendixD}

For supervised learning, it is advantageous to take a narrower test region [white region in Fig.~\ref{fig:fig11}(a)] because more information can be obtained in the training region. 
However, if the test region is too narrow to include the crossing point, the crossing point of the outputs would not be the critical point. Consequently, it is desirable to consider a test region of an appropriate size.

We took the left boundary $\omega=4$ in the main text because this is the value at which the order parameter $n(t)$ decays exponentially, that is, at which the system is in the subcritical region, as shown in Fig.~\ref{fig:fig12}. This result was obtained using the QJMC method. However, the boundary $\omega=8$ was considered because $n(t)$ behaves as it does in the supercritical state. 

To check the sensitivity of the positions of the left and right boundaries, we also considered a test region of $(3\leq \omega \leq 9)$ and then estimated the transition point $\omega_c$ in the thermodynamic limit and the value of the exponent $\nu_\bot$. As shown in Fig.~\ref{fig:fig11}, we obtained the same values of $\omega_c$ and $\nu_\bot$.  
\section{Test of scaling relations using classical Monte Carlo simulations}
\label{appendixE}

We mentioned that $\rho_d(t)$ and $n(t)$ show the same asymptotic behavior, indicating that $\delta=\alpha$ holds~\cite{grassberger1979,landes2012}. In this section, we test this relation using classical Monte Carlo simulations. The models we consider here are the 1D CP and 2D tricritical CP.
Rapidity reversal symmetry reportedly holds for the 1D CP~\cite{hinrichsen2000} and does not hold for the 2D tricritical CP~\cite{lubeck2006,grassberger2006}.
By measuring the slopes in the double-logarithmic plots, we measure the complete set of critical exponents. 

In Fig.~\ref{fig:fig13}(a), the values of all critical exponents for the 1D CP are $z=1.58$, $\delta=0.16$, $\eta=0.31$, $\delta'= 0.16$, and $\alpha=0.16$. Note that $\delta'=\alpha$ and the generalized hyperscaling relation $\eta-D/z=-\delta-\delta'$ hold. In addition, $\delta=\delta'$ because rapidity-reversal symmetry holds. The values of all critical exponents for the 2D tricritical CP [Fig.~\ref{fig:fig13}(b)] are $z=2.11$, $\delta=0.09$, $\eta=-0.35$, and $\delta'= 1.21$. Note that $\delta'=\alpha$ and the generalized hyperscaling relation $\eta-D/z=-\delta-\delta'$ hold. However, $\delta\neq \delta'$ because rapidity-reversal symmetry is broken.

Thus, the scaling relations $\delta = \alpha$ and $\eta-D/z=-\delta-\delta'$, which are thought to be satisfied by the single absorbing-state phase transition, hold; however, $\delta=\delta'$ when rapidity-reversal symmetry holds.
\section{Continuously varying exponent $\alpha$}
\label{appendixF}
\begin{figure}[!t]
\includegraphics[width=1.00\columnwidth]{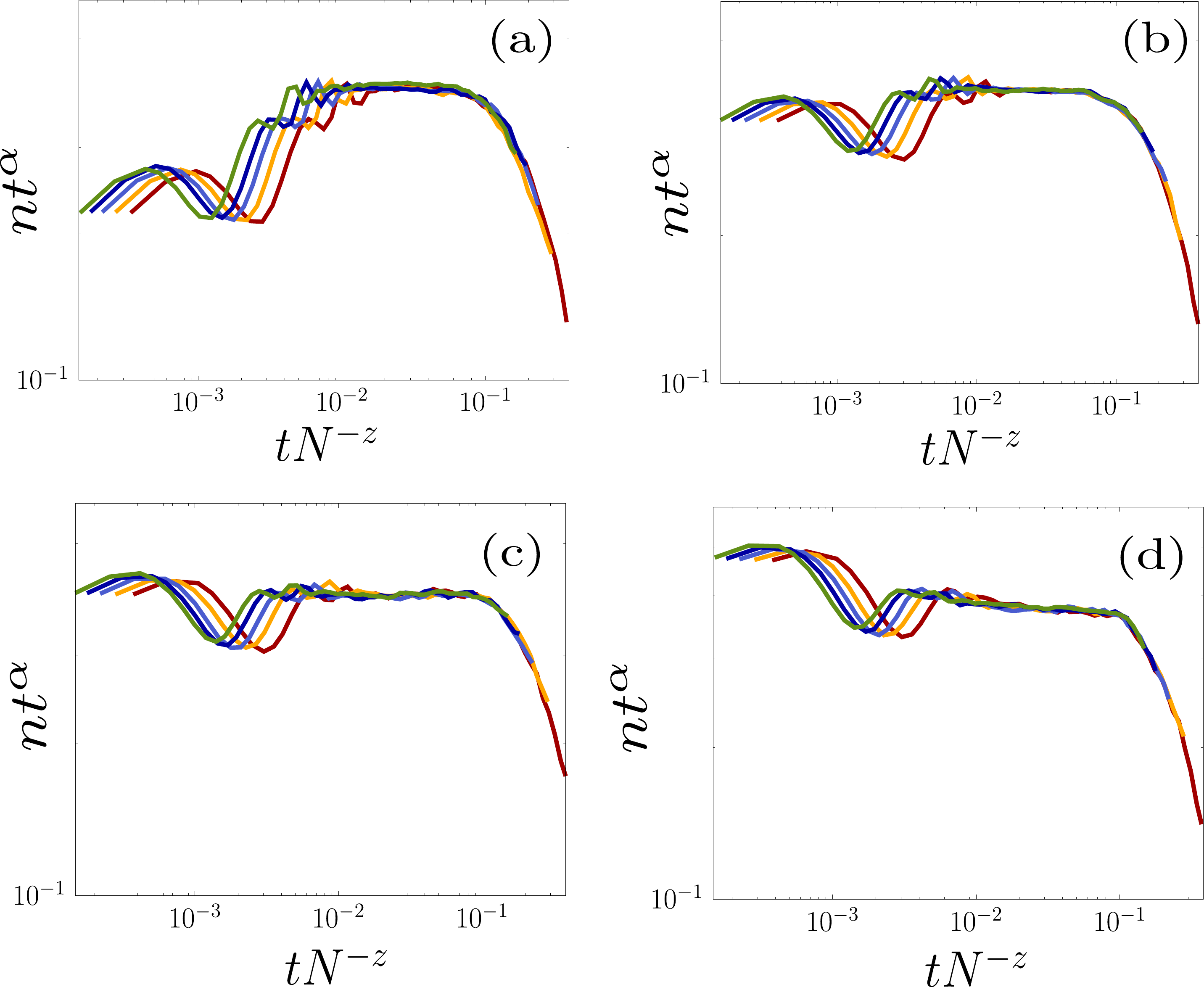}
\caption{Scaling plots of $n(t)t^{\alpha}$ versus $tN^{-z}$ with the classical parameter (a) $\kappa=0.0$, (b) $\kappa=0.2$, (c) $\kappa=0.4$, and (d) $\kappa=0.58$. The critical exponents are taken as $z=1.55$ and $\alpha=0.32$ for (a), $\alpha=0.24$ for (b), $\alpha=0.20$ for (c), and $\alpha=0.16$ for (d).
}
\label{fig:fig14}
\end{figure}
We perform the simulation both QJMC and tensor network approach based on the matrix product state and time evolving block decimation. In our simulation, we set the time steps $dt=0.005$ and the bond dimension $\chi=1024$.
For $\kappa<\kappa_*$, there exists the continuously varying critical exponent $\alpha$, which is different from the DP class. In this section, we shall perform the finite-size scaling for $\kappa\in[0, 0.58]$ in steps of $0.2$ to confirm the continuously varying critical exponent $\alpha$. 
At the transition point, we estimate the exponents $\alpha$ by measuring the slopes directly in the double-logarithmic plots as shown in Fig.~\ref{fig:fig14}(a). The exponent $\alpha$ continuously increases from $0.16$ to $0.32$ as $\kappa$ decreases. Moreover, we confirm those exponents using the data-collapse technique. In Fig.~\ref{fig:fig14}(b-g), we plot $n(t)t^{\alpha}$ versus $tN^{-z}$ for different system sizes $N$ for $\kappa\in[0, 0.58]$ and the critical exponents are listed in Table~\ref{tab:tab2}.
We remark that the critical exponents except $\alpha$ correspond to the DP values within the error-bar for this $\kappa$ region.

\section{Estimation on the critical exponent $\nu_{\|}$ for quantum contact process}\label{appendixG}
\begin{figure}[h!]
\includegraphics[width=1.00\columnwidth]{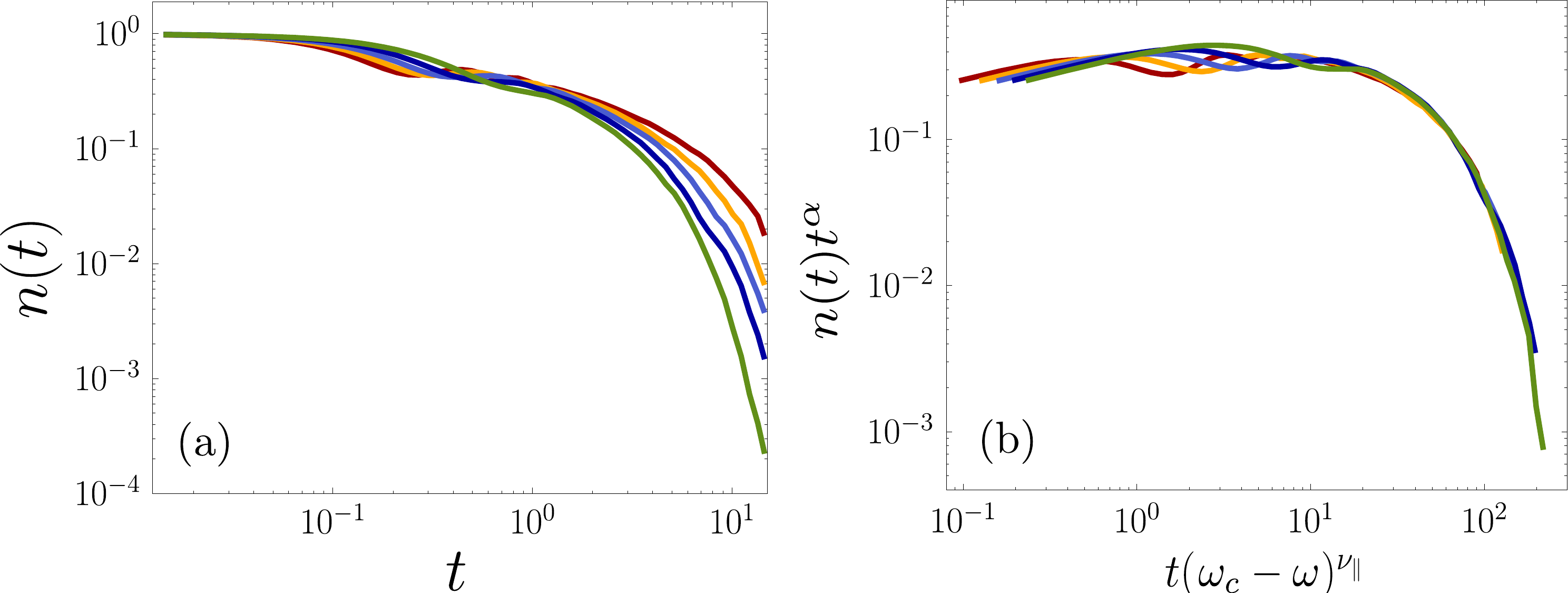}
\caption{Estimates of the critical exponent $\nu_{\|}$ associated with the time correlation. (a) Plot of $n(t)$ as a function of $t$ for different values of $\omega<\omega_c$. 
(b) Data points collapse well onto a single curve for $\omega_c=6.04$, $\alpha=0.32$, and $\nu_{\|}=1.73$. The units of control parameter is given as $\gamma$.} 
\label{fig:fig15}
\end{figure}

We obtained the critical exponent associated with the spatial correlation length $\nu_{\bot}=1.06\pm 0.04$ directly from the NN approach. In this section, we verify the critical exponent associated with the spatial correlation $\nu_{\bot}$ by obtaining the critical exponent associated with the temporal correlation $\nu_{\|}=\nu_{\bot}z$ using QJMC.

In Fig.~\ref{fig:fig15}(b), the exponent $\nu_{\|}$ is obtained from the rescaling plot of $n(t)t^{\alpha}$ versus $t(\omega_c-\omega)^{\nu_{\|}}$ for different $\omega$. The $\omega$ values are taken from the region used in the CCP. We obtain $\nu_{\|}=1.73$ and thus $\nu_{\bot}=\nu_{\|}/z\simeq 1.095$. This value is consistent with that obtained using the NN approach.

\end{appendix}


%

\end{document}